\documentclass[%
 aip,
 amsmath,amssymb,
reprint,%
]{revtex4-2}

\usepackage{graphicx}
\usepackage{dcolumn}
\usepackage{bm}

\usepackage[utf8]{inputenc}
\usepackage[T1]{fontenc}
\usepackage{etoolbox}

\usepackage{xcolor}

\makeatletter
\def\@email#1#2{%
 \endgroup
 \patchcmd{\titleblock@produce}
  {\frontmatter@RRAPformat}
  {\frontmatter@RRAPformat{\produce@RRAP{*#1\href{mailto:#2}{#2}}}\frontmatter@RRAPformat}
  {}{}
}%
\makeatother

\begin{document}

\title{Bath-induced stabilization of classical non-linear response in two dimensional infrared spectroscopy}

\author{Rajesh Dutta}
\thanks{Authors to whom correspondence should be addressed: duttarajesh24@gmail.com}
\author{Mike Reppert}

\affiliation{Purdue University, Department of Chemistry,\\ West Lafayette, IN 47907}

\begin{abstract}
Classical response functions have shown considerable promise in computational 2D IR modeling; however, a simple diagrammatic description, analogous to that for open quantum systems, has been lacking. While a promising diagrammatic approach has recently been introduced for isolated systems, the resulting nonlinear response functions remain unstable at long times, a characteristic feature of integrable classical systems. Here, we extend this framework to incorporate system–bath interactions under the weak-anharmonicity approximation and explore the resulting conditions for bath-induced stabilization. The resulting expression for the weakly anharmonic response function is remarkably simple and exhibits a one-to-one correspondence with the quantum counterpart in the $\hbar\rightarrow0$ limit, offering potential computational advantages in extending the approach to large, multi-oscillator systems. We find that (to lowest order in anharmonicity) the bath-induced stabilization of both linear and nonlinear classical response functions depends sensitively on the nature of spectral density, particularly on the balance between low-frequency and high-frequency components. Application of this classical diagrammatic approach to 2D IR spectroscopy of the amide I band captures the characteristic population-time-dependent dynamics associated with spectral diffusion, suggesting that the approach may prove useful in describing real experimental systems at ambient temperatures.
\end{abstract}


\maketitle

\section{Introduction}
Coherent two-dimensional infrared (2D IR) spectroscopy has attracted growing interest over the past few decades due to its applications in revealing ultrafast dynamics and the structure of biomolecular systems and condensed phases.\cite{hunt20092d,hamm2011concepts,annurev:/content/journals/10.1146/annurev-physchem-040215-112055,10.1063/1.5083966,doi:10.1021/acs.chemrev.9b00813,baiz2024celebrating} Although the standard theory of nonlinear 2D IR spectroscopy is quantum mechanical, it is quite natural to consider its classical analogue, given its widespread use in studying the vibrational motion of large biomolecular systems such as peptides and proteins.\cite{liang2012efficient,10.1063/1.5083966} Moreover, a fully quantum mechanical treatment of such large systems is often computationally prohibitive. Interestingly, classical simulations have also been able to reproduce coherent beating patterns, which are often regarded as signatures of quantum coherence in 2D IR spectra.\cite{10.1063/1.5017985} This raises important questions about the necessity of a purely quantum mechanical description in such cases.

Motivated by these findings, the development of classical analogues to quantum response theory has received increasing attention across different spectroscopic domains.\cite{10.1063/1.5017985,reppert2021diagrammatic,reppert2023equivalence,10.1063/1.5058136,PhysRevA.111.022210} In the case of linear spectroscopy, the classical analogue of quantum linear response theory can be obtained by simply replacing the commutator with a Poisson bracket and taking the $\hbar\rightarrow 0$ limit. Recently, both classical perturbative and non-perturbative theories have been developed in the context of linear electronic spectroscopy.\cite{10.1063/1.5058136,PhysRevA.111.022210} These theories are applicable in the weak-coupling FRET limit and intermediate coupling regimes, including systems such as J or H aggregates, showing promising behavior compared to their quantum mechanical counterparts.

Quantum-classical correspondence is much more complicated, however, in non-linear response due to the need for an accurate treatment of anharmonicity. It is well known that both linear and non-linear response functions in anharmonic systems can exhibit divergence.\cite{van1971case,PhysRevE.53.R1,10.1063/1.1827212,PhysRevLett.95.180405,PhysRevLett.96.030403,kryvohuz2008suppression,10.1063/1.1888485, Wu2001, Noid2004, 10.1063/1.1792211} In particular, classical non-linear response functions for integrable anharmonic systems are generally unstable, with amplitudes diverging along at least one time axis.\cite{PhysRevLett.96.030403} van Kampen long ago pointed out the divergence of the linear response, emphasizing it as a fundamental limitation of response theory, since the response of a nonlinear dynamical system to a weak external field may not remain perturbative at long times.\cite{van1971case} In linear response, thermal or phase-space averaging eliminates this divergence, rendering the long-time response finite, but this is generally not the case for classical nonlinear response.\cite{PhysRevE.53.R1, PhysRevLett.96.030403}  

Although classical nonlinear response theory\cite{kubo1957statistical} was developed long ago alongside its quantum counterpart, it has yet to be fully utilized in practical applications of 2D IR spectroscopy. Computational studies have established that 2D IR response functions can be computed using both classical\cite{la2000third,dellago2003simulation,saito2003off,hasegawa2006calculating,jeon2010direct,sakurai2011does,jeon2014accurate,ito2015notes,10.1063/1.5083966,jung20222d} and semiclassical dynamics.\cite{noid2004semiclassical,alemi2015vibrational,loring2017mean,loring2022calculating} In this context, the formalism developed by Jung and Markland offers a distinct alternative to conventional methods.\cite{jung20222d} Instead of relying on the traditional stability matrix approach, their method reformulates the classical nonlinear response in terms of equilibrium correlation functions, introducing a conceptually different and potentially more efficient computational strategy.

Nevertheless, both classical and semiclassical approaches remain computationally expensive, largely due to the need for extensive trajectory averaging to mitigate dynamical instabilities. More importantly, these methods often lack clear physical interpretability. Because the classical response becomes meaningful only after ensemble averaging, direct insights into the underlying microscopic dynamics are limited.

Efforts have been made to recover interpretability through diagrammatic representations. Mukamel and collaborators proposed a classical perspective in which nonlinear response arises from interference between neighboring classical trajectories.\cite{PhysRevE.53.R1} However, relating these microscopic interference effects to the ensemble-averaged response remains nontrivial. A different approach by Noid and Loring constructs diagrammatic representations in which classical diagrams emerge in the $\hbar\rightarrow0$ limit of a sum over quantum diagrams, with one diagram corresponding to each wavevector-matching condition.\cite{10.1063/1.1792211,10.1063/1.1888485,Noid10112005} Even so, the resulting expressions are algebraically intricate, and generalizing the method to multi-oscillator systems remains a substantial challenge.\cite{10.1063/1.1792211} Recently, a hierarchical equations of motion based approach has been developed in the Wigner-space representation and applied to the 2D IR spectroscopy of water.\cite{hoshino2025analysis} However, the method is computationally demanding and currently feasible only for systems with a very limited number of modes. 

The standard quantum diagrammatic formalism, in contrast, offers a rather transparent analytical framework, where spectral features can be directly related to “arrow-ladder” diagrams representing physical processes such as ground-state bleaching, stimulated emission, or excited-state absorption.\cite{10.1063/1.5083966,Mukamel1995} The clarity and intuitive structure of the quantum diagrammatic expansion underscore its effectiveness and raise an important question: can a corresponding classical diagrammatic framework be developed that preserves both the transparent physical picture of classical physics and the structural simplicity of quantum diagrammatic approach? A promising principle for developing such a classical framework is that the perturbation induced by an electric field on the equilibrium probability density always takes the same form.\cite{10.1063/1.5058136} This enables the system’s response to be described by a discrete set of perturbations to the phase-space probability density, which can, in turn, be mapped to quantum eigenstates.

Based on this principle, a classical diagrammatic approach was recently developed for a single harmonic oscillator, showing term-by-term correspondence with the quantum diagrammatic formulation.\cite{reppert2021diagrammatic} The method relies on a weak-anharmonicity-based perturbative expansion and was applied to an isolated system, i.e., without coupling to an environment or bath degrees of freedom. A damping coefficient or static disorder was introduced \textit{ad hoc} to eliminate the linear divergence arising from anharmonicity. Interestingly, the inclusion of static disorder alone did not fully eliminate the divergence in the non-linear response function. This classical approach based on weak anharmonicity and classical excitations are particularly relevant for modeling the amide I vibrations in proteins and peptides, where the anharmonicity ($\sim16~\text{cm}^{-1}$) is much smaller than either the fundamental vibrational frequency ($\sim1650~\text{cm}^{-1}$) or the thermal energy ($\sim200~\text{cm}^{-1}$) at room temperature. The method was later extended to multi-oscillator systems and demonstrated behavior consistent with quantum mechanical predictions.\cite{reppert2023equivalence}

Despite these advancements, an important open question remains: does the linear divergence persist in classical systems when environmental or bath degrees of freedom are introduced? In this work, our goal is twofold. First, we aim to extend the classical diagrammatic approach by systematic inclusion of bath degrees of freedom and explicit interactions of system and bath, thereby making it more consistent when compared to experiment. Second, we investigate whether a classical bath or fluctuating environment alone can stabilize the response function, or if a dissipative environment is required to yield finite nonlinear responses.

Our analytical results demonstrate that even a classical bath or fluctuating environment can lead to finite nonlinear response functions. However, the outcome critically depends on the form of the spectral density.

The sections of this paper are organized as follows:
Section II introduces the classical Hamiltonian for a single harmonic oscillator and its interaction with the bath in the action-angle representation. Section III presents the formulation of classical excitation, probability density, and propagator. Section IV develops the classical response function and the corresponding diagrammatic approach, along with their application to linear and nonlinear response functions. Section V describes the analogous quantum linear and nonlinear response functions. Section VI discusses the results of classical linear responses for anharmonic oscillator with different spectral densities. Section VII explains 2D correlation spectra and Finally, Section VIII provides the concluding discussion.

\section{\label{sec:level1}Classical Hamiltonian: Action angle representation}

Following previous works\cite{briggs2011equivalence,eisfeld2012classical,mancal2013excitation,10.1063/1.5058136,reppert2020equilibrium,reppert2021diagrammatic,PhysRevA.111.022210}, we consider a classical model consisting of a single harmonic oscillator (the ``system'') coupled to a harmonic bath. The system oscillator is described by its coordinate and momentum, denoted as $q$ and $p$, respectively, while the bath oscillators are characterized by coordinates $Q_k$ and momenta $P_k$. Note here that the system-bath interaction is linear-quadratic rather than bilinear. This form is chosen to mimic the structure of the corresponding quantum Hamiltonian upon canonical quantization.\cite{mancal2013excitation, 10.1063/1.5058136, PhysRevA.111.022210}

\begin{equation}
\begin{split}
\label{eq:classical_Hamiltonian}
  & H_{S}^{(0)}=\frac{1}{2}\left[ \omega _{0}^{2}{{q}^{2}}+{{p}^{2}} \right] \\ 
 & {{H}_{B}}=\frac{1}{2}\sum\limits_{k}{\left[ \Omega _{k}^{2}Q_{k}^{2}+P_{k}^{2} \right]} \\ 
 & {{H}_{SB}}=\sum\limits_{k}{{{\alpha }_{k}}{{\omega }_{0}}{{q}^{2}}{{Q}_{k}}} \\ 
\end{split}
\end{equation}
Here, $\omega_0$ is the (angular) oscillator frequency, and position and momentum coordinates are provided in mass-weighted units.
The dynamics become easier to analyze in the action-angle representation for the system part, as it resembles the discretized structure of quantum systems. The action ($J$) and angle ($\theta$) are defined as  
\begin{equation}
\begin{split}
 & q=\sqrt{\frac{2J}{{{\omega }_{0}}}}\cos \theta  \\ 
 & p=-\sqrt{2{{\omega }_{0}}J}\sin \theta  \\ 
\end{split}  
\end{equation}
The system and system-bath interaction Hamiltonian transforms to
\begin{equation}
\begin{split}
&H_{S}^{(0)}={{\omega }_{0}}J\\
&{{H}_{SB}}=2J{{\cos }^{2}}\theta \sum\limits_{k}{{{\alpha }_{k}}{{Q}_{k}}}\\ 
\end{split}
\end{equation} 
Since we are interested in nonlinear response, it is essential to include the anharmonic part of the Hamiltonian, as the nonlinear response vanishes for purely harmonic systems. The anharmonic part of the system Hamiltonian is as considered as cubic-quartic, given as
\begin{equation}
H_S^{(1)} = \alpha_3 q^3 + \alpha_4 q^4
\end{equation}
Transformation into action-angle representation followed by rotating-wave-approximation leads to only surviving term as follows\cite{reppert2021diagrammatic}
\begin{equation}
H_{S}^{(1)}=\Delta {{J}^{2}}
\end{equation}
where $\Delta$ represent the anharmonicity constant.
\section{\label{sec:level2}Classical excitation, probability density and propagator}

To consider the effect of light matter Liouville operator we define the classical phase space density for the bare system part as follows:\cite{hillery1984distribution,10.1063/1.5017985,reppert2021diagrammatic,reppert2023equivalence}
\begin{equation}
{\varrho }_{ab} = \frac{{{\left( \beta {{\omega }_{0}}J \right)}^{\frac{a+b}{2}}}}{\sqrt{a!b!}}{{e}^{i\left( a-b \right)\theta }}\frac{{{e}^{-\beta {{\omega }_{0}}J}}}{Z} 
\end{equation}
where, Z is the partition function.
Excitation indices $a$ and $b$ can only take non-negative integer values. The classical states are analogous to quantum state with population and coherence. Classical state created by light-matter interaction results in shifting of probability density in phase space. 

Application of the light-matter Liouville operator on the classical phase space density of the system part leads to
\begin{equation}
\begin{aligned}
\hat{L}_{LM}(t) &= \sum_{\alpha} E_{\alpha}(t)\,\hat{l}_{\alpha}, \label{eq:L}\\[0.5em]
\big(\hat{l}_{\alpha}\varrho_{ab}\big)
&= \frac{\mu_{\alpha}^{(1)}}{\sqrt{2\,k_{\mathrm B}T}}
\Big(
    -\sqrt{a}\,\varrho_{a-1,b}
    + \sqrt{b}\,\varrho_{a,b-1} \nonumber \\
&\hspace{8em}
    - \sqrt{a+1}\,\varrho_{a+1,b}
    + \sqrt{b+1}\,\varrho_{a,b+1}
\Big)
\end{aligned}
\end{equation}

Here, $\alpha$ indicates Cartesian coordinate axes. The field independent operator $\hat{l}_\alpha$ obtained by transforming light-matter Hamiltonian in action angle representation. The application of $\hat{l}_\alpha$ changes each state index by $\pm 1$ which is the selection rule for each transition. 

Liouville operators for the system, bath and the interaction between system-bath can be written in the action-angle representation as follows
\begin{equation}
\begin{split}
  & \hat{L}_{S}^{0}=-i{{\omega }_{0}}\frac{\partial }{\partial \theta } \\ 
 & \hat{L}_{S}^{1}=-2i\Delta J\frac{\partial }{\partial \theta } \\ 
 & {{{\hat{L}}}_{SB}}\simeq -i\sum\limits_{k}{{{\alpha }_{k}}{{Q}_{k}}\frac{\partial }{\partial \theta }} \\ 
 & {{{\hat{L}}}_{B}}=-i\sum\limits_{k}{\left[ {{P}_{k}}\frac{\partial }{\partial {{Q}_{k}}}-\Omega _{k}^{2}{{Q}_{k}}\frac{\partial }{\partial {{P}_{k}}} \right]}
 \end{split}
\end{equation}
and the action of system and system-bath interaction Liouville operator to the system part of the classical phase space density provides
\begin{equation}
\begin{split}
  & \hat{L}_{S}^{0}{{\rho }_{ab}}=\left( a-b \right){{\omega }_{0}}{{\varrho }_{ab}} \\ 
 & \hat{L}_{S}^{1}{{\rho }_{ab}}=2\Delta J\left( a-b \right){{\varrho }_{ab}} \\ 
 & {{{\hat{L}}}_{SB}}{{\rho }_{ab}}=\sum\limits_{k}{\left( a-b \right){{\alpha }_{k}}{{Q}_{k}}{{\varrho }_{ab}}}
\end{split}  
\end{equation}
Initially, prior to being excited electromagnetically the system and bath remain in a factorized equilibrium state:
\begin{equation}
{{\rho }_{eq}}={{\rho }_{S}^{eq}}\otimes {{\rho }_{B}^{eq}} .
\end{equation}
Here, the equilibrium system state corresponds to setting both excitation indices $a$ and $b$ to zero, i.e., $\rho_S^\text{eq} = \varrho_{00}$.

\section{\label{sec:level1}Classical response function: Diagrammatic expansion}

Similar to the quantum case, the classical light matter interaction can be expressed using a perturbative expansion. The $n$th order response function can be written in terms of delay times ($\tau_n$) or the interval between light matter excitation as follows\cite{reppert2023equivalence, reppert2021diagrammatic}
\begin{equation}
\scalebox{0.84}{$
R^{(n)}_{\alpha_{n+1} \cdots \alpha_1}(\tau_n, \ldots, \tau_1) 
= (-i)^n \left\langle 
\mu_{\alpha_{n+1}} e^{-i \hat{L}_{\text{tot}} \tau_n} 
\hat{\ell}_{\alpha_n} \cdots 
e^{-i \hat{L}_{\text{tot}} \tau_1} \hat{\ell}_{\alpha_1} 
\right\rangle_{\mathrm{eq}}
$}
\end{equation}
In the $n$th-order response tensor, the first $n$ of the dipole components originates due to the light-matter interaction, whereas the last dipole component represents the induced polarization of the matter.
Here, the total Liouvillian corresponds to 
\begin{equation}
{{\hat{L}}_{tot}}=\hat{L}_{S}^{(0)}+\hat{L}_{SB}^{{}}+\hat{L}_{B}^{{}}+\hat{L}_{S}^{(1)}  
\end{equation}
and where the notation $\left<...\right>_{eq}$ denotes the average over system and bath coordinates.
\begin{equation}
\left<...\right>_{eq}=\prod\limits_{k} \int\limits_{-\infty}^{\infty} dP_k \int\limits_{-\infty}^{\infty} dQ_k 
\int\limits_{0}^{\infty} dJ \int\limits_{0}^{2\pi} d\theta...\rho_{eq} 
\end{equation}
We note that all non-linear responses ($n>1$) vanish for a harmonic oscillator, indicating that the dynamical properties of a harmonic system can be fully captured by the linear response function. The vanishing of the non-linear response for a harmonic oscillator was explained using arrow-ladder diagrammatic expansion.

For a small anharmonicity ($\Delta$), the response function can be expanded perturbatively in the increasing powers of $\Delta$. Here, we are interested in the first-order perturbation due to the anharmonicity. First order perturbation with respect to the anharmonicity expands the system part of the initial probability density and the propagator which leads to the time dependency of the anharmonic part of the Liouvillian.

As the anharmonic Liouvillian commutes with the exponential operator, the effect of the time dependent anharmonic Liouvillian on the system part of the probability density takes the form 
\begin{equation}
\begin{split}
\hat{L}_{1}(t) \varrho_{ab} &= 2t \Delta J(a - b) \varrho_{ab} \\
&\equiv \frac{2t \Delta(a - b) }{\beta\omega_0}
\sqrt{(a + 1)(b + 1)} \, \varrho_{a+1, b+1}
\end{split}
\end{equation}
Thus the first order perturbation introduces a linear time dependence and increases both excitation indices of the
input state by one.
Under first order perturbation, only one term of the response function survives the ensemble average, which (for $n>1$) can be written as\cite{reppert2021diagrammatic} 
\begin{widetext}
\begin{equation}
R_{\alpha_{n+1} \cdots \alpha_1}^{(n)}(\tau_n, \ldots, \tau_1) = 
(-i)^{n+1} \left\langle \mu_{\alpha_{n+1}} 
e^{-i \hat{L}_h^{(0)} \tau_n} \hat{L}_1(\tau_n) \hat{l}_{\alpha_n} 
e^{-i \hat{L}_h^{(0)} \tau_{n-1}} \hat{l}_{\alpha_{n-1}} \cdots 
e^{-i \hat{L}_h^{(0)} \tau_1} \hat{l}_{\alpha_1} \right\rangle_{\mathrm{eq}}
\end{equation}
\end{widetext}
where ${{\hat{L}}_{h}}=\hat{L}_{S}^{(0)}+\hat{L}_{SB}^{{}}+\hat{L}_{B}^{{}}$.
The vanishing of other term involving $\hat{L}_1(\tau_{n-1})$ can be explained using the diagrammatic approach described elsewhere.\cite{reppert2021diagrammatic} The derivation of Ref.~\onlinecite{reppert2021diagrammatic} for the bare system response holds in the present case as well (i.e., with system-bath interactions) since the only difference in the two derivations is the bath-induced phase factor $f_\text{prop}^{SB(\xi,i)}$ of Eq.\eqref{eq:fprop}, which has no effect on the diagrammatic cancellations. Here, we provided only the non-linear response function in the form of different propagator in diagrammatic representation as follows (see Ref.~\onlinecite{reppert2021diagrammatic} for pictures of the diagrams)
\begin{widetext}
\begin{equation}
R_{{{\alpha }_{n+1}}...{{\alpha }_{1}}}^{\left( n \right)}\left( {{\tau }_{n}},...,{{\tau }_{1}}\right)={{\left( -i \right)}^{n+1}}\sum\limits_{\xi }{f_{\left( anh \right)}^{\left( \xi  \right)}\left( {{\tau }_{3}} \right)}{{f}_{rad}}\left( N_{fin}^{\left( \xi  \right)}+1 \right)\prod\limits_{i=1}^{n}{f_{trans}^{\left( \xi ,i \right)}f_{prop}^{S\left( \xi ,i \right)}\left( {{\tau }_{i}} \right)f_{prop}^{SB\left( \xi ,i \right)}\left( {{\tau }_{i}} \right)}
\end{equation}
\end{widetext}
where 
\begin{equation}
f_{trans}^{\left( \xi ,i \right)}\equiv \pm \mu _{{{\alpha }_{i}}}^{\left( 1 \right)}\sqrt{\frac{N_{\xi }^{\left( i \right)}}{2{{k}_{B}}T}}  .
\end{equation}
The index $\xi $ refers to a particular $n$-rung ladder diagram and $N_{\xi }^{\left( i \right)}$ is the larger of the indices involved in the transition across rung $i$. The sign is positive for transitions on the right and negative for transitions on the left. An even number of interactions on the left thus corresponds to an overall positive sign on the ladder, while an odd number of interactions on the left generate a negative sign. The light-matter interaction factor is
\begin{equation}
{{f}_{rad}}\left( N_{fin}^{\left( \xi  \right)}+1 \right)=\frac{\mu _{{{\alpha }_{n+1}}}^{\left( 1 \right)}}{{{\omega }_{0}}}\sqrt{\frac{\left( N_{fin}^{\left( \xi  \right)}+1 \right){{k}_{B}}T}{2}}
\end{equation}
where $N_{fin}^{\left( \xi  \right)}=\max \left( a_{\xi }^{\left( n \right)},b_{\xi }^{\left( n \right)} \right)$, and $a_{\xi }^{\left( n \right)}$ and  $b_{\xi }^{\left( n \right)}$ are the left and right indices above the $n$th rung before the anharmonic contribution comes in.
Sign of the anharmonic contribution is positive if the left index of the final state is larger than the right and vice versa.
The contribution for the propagation in the interval  for the system part follows
\begin{equation}
f_{prop}^{S\left( \xi ,i \right)}\left( {{\tau }_{i}} \right)\equiv \exp \left[ -i\left( a_{\xi }^{\left( i \right)}-b_{\xi }^{\left( i \right)} \right){{\omega }_{0}}{{\tau }_{i}} \right]    
\end{equation}
From the system-bath interaction and bath propagation results the following factor:
\begin{widetext}
\begin{equation}
\begin{split}
f_{\text{prop}}^{SB(\xi, i)}(\tau_i) 
&\equiv \exp\left[ i \sum_{k} \frac{\alpha_k}{\Omega_k^2} \left(P_k - P_k(\tau)\right) 
\left(a_{\xi}^{(i)} - b_{\xi}^{(i)}\right) \right] e^{-i \hat{L}_B \tau} \\
&= \exp\left\{ i \left[ \sum_{k} \frac{\alpha_k}{\Omega_k^2} \left(1 - \cos(\Omega_k \tau)\right) P_k 
+ \sum_{k} \frac{\alpha_k}{\Omega_k} \sin(\Omega_k \tau) Q_k \right] \right\} 
\left(a_{\xi}^{(i)} - b_{\xi}^{(i)}\right) e^{-i \hat{L}_B \tau}
\end{split}
\label{eq:fprop}
\end{equation}  
\end{widetext}
The detailed derivation of this propagator is given in the APPENDIX A.

\subsection{Classical linear response}

To first order in anharmonicity, the linear response function can be expanded as
\begin{widetext}
\begin{equation}
R^1(\tau) = -i \left\langle \mu_{\alpha_2} e^{-i\left( \hat{L}_{S}^{(0)} + \hat{L}_{SB} + \hat{L}_{B} \right)\tau} \hat{l}_{\alpha_1} \right\rangle_{\text{eq}} 
+ (-i)^2 \left\langle \mu_{\alpha_2} e^{-i\left( \hat{L}_{S}^{(0)} + \hat{L}_{SB} + \hat{L}_{B} \right)\tau} \hat{L}_1(\tau_1) \hat{l}_{\alpha_1} \right\rangle_{\text{eq}}
\end{equation}
\end{widetext}
or after evaluating the bath average
\begin{equation}
\begin{aligned}
R^{1}(\tau) =\ & \mu_{\alpha_1}^{(1)} \mu_{\alpha_2}^{(1)} \left[ 
\frac{\sin(\omega_0 \tau)}{\omega_0} 
+ \frac{4 \Delta \tau k_B T \cos(\omega_0 \tau)}{\omega_0^2} 
\right] \\
& \times \exp \left[ 
- \int_0^{\infty} d\Omega \, 
\frac{2 \Lambda(\Omega) \left( 1 - \cos(\Omega \tau) \right)}{\Omega^3 \beta} 
\right]
\label{eq:linear_response}
\end{aligned}
\end{equation}

where the classical bath spectral density is defined as
\begin{equation}
\Lambda\left( \Omega  \right)=\frac{1}{2}\sum\limits_{k}{\frac{\alpha _{k}^{2}}{\Omega _{k}}\delta \left( \Omega -{{\Omega }_{k}} \right)}
\end{equation}
The relation between the quantum ($C''$) and classical ($\Lambda$) spectral density can be written as
\begin{equation}
C''\left(\Omega\right)=\hbar\Lambda\left(\Omega\right)
\end{equation}
First, we assume a Drude-Lorentz spectral density
\begin{equation}
\label{eq:drude_lorenz_classical}
\Lambda \left( \Omega  \right)=\frac{2\Lambda_{0} \Omega \gamma }{\pi\left({{\Omega }^{2}}+{{\gamma }^{2}}\right)}
\end{equation}
The response function becomes 
\begin{equation}
\begin{aligned}
\label{eq:drude_lorenz_response}
R^{1}(\tau) &= \mu_{\alpha_1}^{(1)} \mu_{\alpha_2}^{(1)} 
\left[ 
\frac{\sin(\omega_0 \tau)}{\omega_0} 
+ \frac{4 \Delta \tau k_B T \cos(\omega_0 \tau)}{\omega_0^2} 
\right] \\
&\quad \times \exp\left[
-\frac{2 \Lambda_0}{\beta \gamma} 
\left( 
\tau - \frac{1 - e^{-\gamma \tau}}{\gamma} 
\right)
\right]
\end{aligned}
\end{equation}
In the limit of very fast bath or homogeneous broadening limit $\gamma\rightarrow\infty$ we obtain exponential decay or Lorentzian behavior in frequency domain. For a extremely slow bath $\gamma\rightarrow 0$, the response function follows a gaussian decay i.e. we recover inhomogeneous broadening due to static disorder or gaussian behavior in frequency domain. We also note that even in the presence of anharmonicty the linear response function does not diverge in the long time limit for Drude-Lorentz spectral density.

Next, we derive linear response to verify its stability for generalized spectral density with exponential cut-off as

\begin{equation}
\label{eq:gen_spec_den}
\Lambda_n(\Omega) = \frac{A_n}{n!} \frac{\Omega^n}{\Omega_c^{n-1}} e^{-\Omega/\Omega_c}
\end{equation}
where $A_n$ determines the strength of the coupling with unit of inverse of $\hbar$, and $\Omega_c$ is the cutoff frequency.
\begin{figure}[h]
 \centering
 \includegraphics[height=6.3cm]{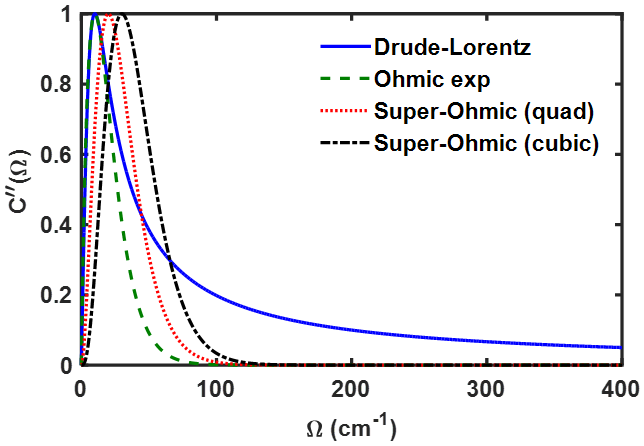}
\caption{Different spectral densities with reorganization energy 2 {cm}$^{-1}$ and cut-off frequency 10 {cm}$^{-1}$. Spectral densities are normalized to unity with respect to the individual maxima. }
 \label{fig:fig1}
\end{figure}
In Fig.~\ref{fig:fig1}, we plotted different spectral densities with the same reorganization energy of 2 cm$^{-1}$. The reorganization energy is defined as
\begin{equation}
    E_r=\hbar\int_0^{\infty} d\Omega \frac{C''(\Omega)}{\Omega}
\end{equation}
Fig.~\ref{fig:fig1} exhibits a prolonged high-frequency tail for the Drude–Lorentz spectral density, which is absent in the others. Moreover, as the spectral density changes from Ohmic to super-Ohmic, it broadens and its maximum shifts toward the higher-frequency regime.

For Ohmic spectral density with $n=1$, the linear response follows
\begin{equation}
\begin{split}
\label{eq:ohmic_response}
R^{1}(\tau) &= \mu_{\alpha_1}^{(1)} \mu_{\alpha_2}^{(1)} \Bigg[ 
\frac{\sin(\omega_0 \tau)}{\omega_0} 
+ \frac{4 \Delta \tau k_B T \cos(\omega_0 \tau)}{\omega_0^2} 
\Bigg] \\
&\quad \times \exp\left[ 
-\frac{A_1}{\beta} \left( 
\tau \tan^{-1}(\Omega_c \tau) 
- \frac{1}{2\Omega_c} \ln(1 + \Omega_c^2 \tau^2) 
\right) 
\right]
\end{split}
\end{equation}

Thus the stability of the linear response at long times is also seen for the Ohmic spectral density with an exponential cut-off. In this case, we observe behavior similar to homogeneous or inhomogeneous broadening, depending on whether the cut-off value is large or small, as in the Drude–Lorentz spectral density.
For the super-Ohmic spectral density, the response function also remains stable (i.e., it tends toward zero as $\tau \to \infty$) when $n=2$:
\begin{equation}
\begin{split}
\label{eq:super_ohmic_n_2_response}
R^{1}(\tau) 
&= \mu_{\alpha_1}^{(1)} \mu_{\alpha_2}^{(1)} 
\Bigg[ 
\frac{\sin(\omega_0 \tau)}{\omega_0} 
+ \frac{4 \Delta \tau k_B T \cos(\omega_0 \tau)}{\omega_0^2} 
\Bigg] \\
&\quad \times \left(1 + \Omega_c^2 \tau^2 \right)^{-\frac{A_2}{2 \beta \Omega_c}}
\end{split}
\end{equation}
For $n=3$, however, the response function becomes unstable in the long time limit
\begin{equation}
\begin{split}
\label{eq:super_ohmic_n_3_response}
R^{1}(\tau) 
&= \mu_{\alpha_1}^{(1)} \mu_{\alpha_2}^{(1)} 
\Bigg[ 
\frac{\sin(\omega_0 \tau)}{\omega_0} 
+ \frac{4 \Delta \tau k_B T \cos(\omega_0 \tau)}{\omega_0^2} 
\Bigg] \\
&\quad \times \exp\left[ 
-\frac{A_3}{3 \beta \Omega_c^2} 
\cdot \frac{\Omega_c^3 \tau^2}{1 + \Omega_c^2 \tau^2} 
\right]
\end{split}
\end{equation}
since the exponential term on the second line tends toward a constant when $\tau \to \infty$. \\

\subsection{Classical non-linear response}

In this section, using the diagrammatic representation, we calculate the third-order non-linear response functions relevant to coherent two-dimensional IR spectroscopy. The diagrams can be categorized into three non-rephasing, three rephasing, and two double quantum coherence diagrams along with their complex conjugates (mirror images). In the non-rephasing case, phase evolution during $\tau_1$ and $\tau_3$ teakes place with the same sign, whereas in the rephasing case, the signs are opposite. In the case of double quantum coherence, the propagation during the delay time $\tau_2$ evolves with twice the transition frequency $\omega_0$. For brevity, we refrain from including the diagrams here, as they have already been provided and explained elsewhere.\cite{reppert2021diagrammatic}

Here, we provide only the expressions of non-rephasing and rephasing response functions, as the 2D correlation plot requires the Fourier transform of these two. The use of the evolution equation for the bath co-ordinates ($Q_k$ and $P_k$) and the definition of bath spectral density leads to the following expression for non-rephasing and rephasing response 
\begin{widetext}
\begin{equation}
\begin{aligned}
R_{\alpha_1 \alpha_2 \alpha_3 \alpha_4}^{(NR)}(\tau_1, \tau_2, \tau_3) 
&= \frac{\mu_{\alpha_1}^{(1)} \mu_{\alpha_2}^{(1)} \mu_{\alpha_3}^{(1)} \mu_{\alpha_4}^{(1)} \Delta \tau_3}{\omega_0^2} 
\, e^{-i \omega_0 (\tau_1 + \tau_3)} \\
&\quad \times \left\langle 
    \exp\left[i \sum_{k} \frac{\alpha_k}{\Omega_k^2} \left(P_k - P_k(\tau_3)\right)\right] 
    e^{-i\hat{L}_B \tau_3} 
    e^{-i\hat{L}_B \tau_2} 
    \exp\left[i \sum_{k} \frac{\alpha_k}{\Omega_k^2} \left(P_k - P_k(\tau_1)\right)\right] 
    e^{-i\hat{L}_B \tau_1} 
\right\rangle_B + \text{c.c.} \\
&= \frac{\mu_{\alpha_1}^{(1)} \mu_{\alpha_2}^{(1)} \mu_{\alpha_3}^{(1)} \mu_{\alpha_4}^{(1)} \Delta \tau_3}{\omega_0^2} 
\, e^{-i \omega_0 (\tau_1 + \tau_3)} \\
&\quad \times \exp\bigg[ 
  -\int_0^\infty d\Omega \frac{2\Lambda(\Omega)}{\beta \Omega^3} \left(
    2 - \cos(\Omega \tau_1) - \cos(\Omega \tau_2) - \cos(\Omega \tau_3) 
    + \cos(\Omega (\tau_1 + \tau_2)) \right. \\
&\qquad \left.
    + \cos(\Omega (\tau_2 + \tau_3)) 
    - \cos(\Omega (\tau_1 + \tau_2 + \tau_3))
  \right) 
\bigg] + \text{c.c.}
\end{aligned}
\end{equation}    
\end{widetext}
\begin{widetext}
\begin{equation}
\begin{aligned}
R_{\alpha_1 \alpha_2 \alpha_3 \alpha_4}^{(R)}(\tau_1, \tau_2, \tau_3) 
&= \frac{\mu_{\alpha_1}^{(1)} \mu_{\alpha_2}^{(1)} \mu_{\alpha_3}^{(1)} \mu_{\alpha_4}^{(1)} \Delta \tau_3}{\omega_0^2} 
\, e^{-i \omega_0 (\tau_3 - \tau_1)} \\
&\quad \times \left\langle 
    \exp\left[i \sum_{k} \frac{\alpha_k}{\Omega_k^2} \left(P_k - P_k(\tau_3)\right)\right] 
    e^{-i\hat{L}_B \tau_3} 
    e^{-i\hat{L}_B \tau_2} 
    \exp\left[-i \sum_{k} \frac{\alpha_k}{\Omega_k^2} \left(P_k - P_k(\tau_1)\right)\right] 
    e^{-i\hat{L}_B \tau_1} 
\right\rangle_B + \text{c.c.} \\
&= \frac{\mu_{\alpha_1}^{(1)} \mu_{\alpha_2}^{(1)} \mu_{\alpha_3}^{(1)} \mu_{\alpha_4}^{(1)} \Delta \tau_3}{\omega_0^2} 
\, e^{-i \omega_0 (\tau_3 - \tau_1)} \\
&\quad \times \exp\bigg[ 
  -\int_0^\infty d\Omega \frac{2\Lambda(\Omega)}{\beta \Omega^3} \left(
    2 - \cos(\Omega \tau_1) + \cos(\Omega \tau_2) - \cos(\Omega \tau_3) 
    - \cos(\Omega (\tau_1 + \tau_2)) \right. \\
&\qquad \left.
    - \cos(\Omega (\tau_2 + \tau_3)) 
    + \cos(\Omega (\tau_1 + \tau_2 + \tau_3))
  \right) 
\bigg] + \text{c.c.}
\end{aligned}
\end{equation}
\end{widetext}
Temporal divergence is a characteristic feature of the nonlinear response in integrable classical systems, irrespective of the strength of the anharmonicity. However, coupling to a bath or performing thermal averaging can lead to stabilization of the response function. The stability of the response function is examined here using Drude–Lorentz, Ohmic, and super-Ohmic spectral densities with exponential cut-off.
For Drude-Lorentz spectral density the non-rephasing response becomes

\begin{widetext}
\begin{equation}
\begin{aligned}
R_{\alpha_1 \alpha_2 \alpha_3 \alpha_4}^{(NR)}(\tau_1, \tau_2, \tau_3) &= \frac{\mu_{\alpha_1}^{(1)} \mu_{\alpha_2}^{(1)} \mu_{\alpha_3}^{(1)} \mu_{\alpha_4}^{(1)} \Delta \tau_3}{\omega_0^2} 
\, e^{-i \omega_0 (\tau_1 + \tau_3)} \\
&\quad \exp\Big[-\frac{2 \Lambda_0}{\beta \gamma}  \Big( 
\left( \tau_1 - \frac{1 - e^{-\gamma \tau_1}}{\gamma} \right)
+ \left( \tau_2 - \frac{1 - e^{-\gamma \tau_2}}{\gamma} \right)
+ \left( \tau_3 - \frac{1 - e^{-\gamma \tau_3}}{\gamma} \right) \\
&\quad 
- \left( \left[\tau_1+\tau_2\right] - \frac{1 - e^{-\gamma \left[\tau_1+\tau_2\right]}}{\gamma} \right) 
- \left( \left[\tau_2+\tau_3\right] - \frac{1 - e^{-\gamma \left[\tau_2+\tau_3\right]}}{\gamma} \right) \\
&\quad 
+ \left( \left[\tau_1+\tau_2+\tau_3\right] - \frac{1 - e^{-\gamma \left[\tau_1+\tau_2+\tau_3\right]}}{\gamma} \right) 
\Big) \Big]+\text{c.c.}
\end{aligned}
\end{equation}
\end{widetext}

At, the limit of very fast bath or homogeneous broadening limit $\gamma\rightarrow\infty$, the non-rephasing response function becomes,
\begin{equation}
\begin{split}
R_{\alpha_1 \alpha_2 \alpha_3 \alpha_4}^{(NR)}(\tau_1, \tau_2, \tau_3) 
= &\; \frac{\mu_{\alpha_1}^{(1)} \mu_{\alpha_2}^{(1)} \mu_{\alpha_3}^{(1)} \mu_{\alpha_4}^{(1)} \Delta \tau_3}{\omega_0^2}e^{-i \omega_0 (\tau_1 + \tau_3)}  \\
&\times 
\exp\left[-\frac{2 \Lambda_0}{\beta \gamma} (\tau_1 + \tau_3)\right]+ \text{c.c.}
\end{split}
\end{equation}

At, the limit of slow bath or inhomogeneous broadening limit $\gamma\rightarrow 0$, the non-rephasing response function becomes,
\begin{equation}
\begin{split}
R_{\alpha_1 \alpha_2 \alpha_3 \alpha_4}^{(NR)}(\tau_1, \tau_2, \tau_3) 
= &\; \frac{
\mu_{\alpha_1}^{(1)} \mu_{\alpha_2}^{(1)} 
\mu_{\alpha_3}^{(1)} \mu_{\alpha_4}^{(1)} 
\Delta \tau_3}{\omega_0^2}e^{-i \omega_0 (\tau_1 + \tau_3)}  \\
&\times 
\exp\left[ -\frac{\Lambda_0}{\beta} (\tau_1 + \tau_3)^2 \right] + \text{c.c.}
\end{split}
\end{equation}

Hence, for Drude-Lorentz spectral density, non-linear response function is stable and temporal divergence vanishes.

For the Drude-Lorentz spectral density the rephasing response can be written as
\begin{widetext}
\begin{equation}
 \begin{aligned}
R_{\alpha_1 \alpha_2 \alpha_3 \alpha_4}^{(R)}(\tau_1, \tau_2, \tau_3) &= \frac{\mu_{\alpha_1}^{(1)} \mu_{\alpha_2}^{(1)} \mu_{\alpha_3}^{(1)} \mu_{\alpha_4}^{(1)} \Delta \tau_3}{\omega_0^2} 
\, e^{-i \omega_0 (\tau_3 - \tau_1)} \\
& \quad \exp\Big[-\frac{2 \Lambda_0}{\beta \gamma}  \Big( 
\left( \tau_1 - \frac{1 - e^{-\gamma \tau_1}}{\gamma} \right)
    - \left( \tau_2 - \frac{1 - e^{-\gamma \tau_2}}{\gamma} \right)
    + \left( \tau_3 - \frac{1 - e^{-\gamma \tau_3}}{\gamma} \right) \\
&\quad 
    + \left( \left[\tau_1+\tau_2\right] - \frac{1 - e^{-\gamma \left[\tau_1+\tau_2\right]}}{\gamma} \right) 
    + \left( \left[\tau_2+\tau_3\right] - \frac{1 - e^{-\gamma \left[\tau_2+\tau_3\right]}}{\gamma} \right) \\
&\quad 
    - \left( \left[\tau_1+\tau_2+\tau_3\right] - \frac{1 - e^{-\gamma \left[\tau_1+\tau_2+\tau_3\right]}}{\gamma} \right) 
\Big) \Big] +\text{c.c.}
\end{aligned}   
\end{equation}   
\end{widetext}
At the limit of very fast bath or homogeneous broadening limit $\gamma\rightarrow\infty$, the rephasing response function becomes,
\begin{equation}
\begin{split}
R_{\alpha_1 \alpha_2 \alpha_3 \alpha_4}^{(R)}(\tau_1, \tau_2, \tau_3) 
= &\; \frac{
\mu_{\alpha_1}^{(1)} \mu_{\alpha_2}^{(1)} 
\mu_{\alpha_3}^{(1)} \mu_{\alpha_4}^{(1)} 
\Delta \tau_3}{\omega_0^2} e^{-i \omega_0 (\tau_3 - \tau_1)}\\
&\times  
\exp\left[ -\frac{2 \Lambda_0}{\beta \gamma} (\tau_1 + \tau_3) \right] + \text{c.c.}
\end{split}
\end{equation}
At the limit of slow bath or inhomogeneous broadening limit $\gamma\rightarrow 0$, the rephasing response function becomes,
\begin{equation}
\begin{split}
R_{\alpha_1 \alpha_2 \alpha_3 \alpha_4}^{(R)}(\tau_1, \tau_2, \tau_3) 
= &\; \frac{
\mu_{\alpha_1}^{(1)} \mu_{\alpha_2}^{(1)} 
\mu_{\alpha_3}^{(1)} \mu_{\alpha_4}^{(1)} 
\Delta \tau_3}{\omega_0^2}e^{-i \omega_0 (\tau_3 - \tau_1)}  \\
&\times 
\exp\left[ -\frac{\Lambda_0}{\beta} (\tau_3 - \tau_1)^2 \right] + \text{c.c.}
\end{split}
\end{equation}

Hence, for Drude-Lorentz spectral density, rephasing spectra is stable except at the static disorder or inhomogeneous broadening limit when $\tau_1=\tau_3$.

Expression of non-linear response functions for generalized spectral densities (Eq.~\eqref{eq:gen_spec_den}) are provided in APPENDIX B. Similarly to the Drude-Lorentz spectral density, for the Ohmic spectral density with exponential cutoff, the nonrephasing response exhibits an exponential decay in the large cutoff limit (homogeneous broadening) and a Gaussian decay in the small cutoff limit (inhomogeneous broadening). The rephasing response also shows exponential decay in the large cut-off limit. However, in the small cut-off or inhomogeneous broadening limit, the rephasing response displays a divergence along the $\tau_1 = \tau_3$ axis.

For super-Ohmic spectral densities ($n=2$), both the non-rephasing and rephasing response functions remain stable in the long-time limit, as the response function includes a decaying component.  
However, for a super-Ohmic spectral density with $n=3$, a linear divergence caused by anharmonicity persists, as both the non-rephasing and rephasing response functions become independent of the decay component in the long-time limit.

\section{Quantum Response Function}
Upon canonical quantization, the total Hamiltonian becomes
\begin{equation}
\begin{aligned}
\hat{H}_{\text{tot}}^a &= \hat{H}_S + \hat{H}_{SB} + \hat{H}_B \nonumber \\
&= \sum_a |a\rangle E_a \langle a| 
+ \sum_a |a\rangle \sum_k \hbar\alpha_k^a \hat{Q}_k \langle a| \\
&\quad + \frac{1}{2} \sum_k \left(\hat{P}_k^2 
+ \Omega_k^2 \hat{Q}_k^2\right)
\end{aligned}
\end{equation}

The anharmonic contribution to the energy levels is included in the $E_a$ term. Under first-order perturbation, the overall system part of the energy can be written as
\begin{equation}
 E_a=\hbar\omega_{0}\left( a+\frac{1}{2} \right)+\hbar^2 \Delta \left( {{a}^{2}}+a+\frac{1}{4} \right) \end{equation}
The anharmonic transition frequency $(E_{a+1}-E_a)/\hbar$ is given as
\begin{equation}
\begin{split}
\omega_{10} &\approx \omega_0 + 2\hbar \Delta \\
\omega_{21} &\approx \omega_0 + 4\hbar \Delta
\end{split}
\end{equation}
The system bath coupling term is diagonal with respect to the system eigenstates describing
the fluctuation of the system energy-levels due to the
system–bath interaction. As, we are interested in pure dephasing model, we neglect off-diagonal system bath coupling term responsible for bath induced transition between eigen states of the system.  
\subsection{Quantum linear response}
The linear response is defined as
\begin{equation}
{{R}^{1}}\left( \tau  \right)=\frac{i}{\hbar }\left\langle \left[ \bm{\mu} \left( \tau  \right),\bm{\mu} \left( 0 \right) \right] \right\rangle
\end{equation}
To evaluate the dipole dipole correlation we used the interaction representation for the time-dependent fluctuation and followed by using time ordered cumulant expansion, the corresponding response function can be written as
\begin{equation}
\begin{aligned}
R^{1}(\tau) 
&= \frac{i}{\hbar} 
   \mu_{\alpha_1}^{(01)} \mu_{\alpha_2}^{(10)} 
   \Bigg[ \left( e^{-i \omega_{10}\tau} - \text{c.c.} \right) \nonumber \\
&\quad \times \exp \Bigg\{ 
   - \int_0^{\infty} 
   \frac{C''(\Omega)}{\Omega^2} 
   \Bigg[ 
      \left( 1 - \cos(\Omega \tau) \right) 
      \coth\left( \frac{\beta \hbar \Omega}{2} \right) \nonumber \\
&\qquad\qquad + i \left( \sin(\Omega \tau) - \Omega \tau \right) 
   \Bigg] 
   d\Omega 
   \Bigg\}
   \Bigg]
\end{aligned}
\end{equation}
Here, the quantum spectral density is defined as $C''(\Omega)=\hbar\Lambda(\Omega)$. The Harmonic dipole moments are unperturbed to the first order of anharmonicity and can be expressed as
\begin{equation}
\mu_{\alpha}^{(a, a+1)} = \mu_{\alpha}^{(a+1, a)} \approx \mu_{\alpha}^{(1)} \sqrt{\frac{\hbar (a + 1)}{2 \omega_0}}
\end{equation}
The overall quantum linear response function to the first order of anharmonicity can be written as
\begin{equation}
\begin{aligned}
R^{1}(\tau) 
&= \frac{\mu_{\alpha_1}^{(1)} \mu_{\alpha_2}^{(1)}}{\omega_0} 
   \Bigg[ 
     \sin(\omega_0 \tau) 
     + 2\hbar \Delta \tau \cos(\omega_0 \tau) 
   \Bigg] \\
&\quad \times \exp \Bigg\{ 
  - \int_0^{\infty} 
    \frac{C''(\Omega)}{\Omega^2} 
    \Bigg[ 
      \left(1 - \cos(\Omega \tau)\right) 
      \coth\left( \frac{\beta \hbar \Omega}{2} \right) \\
&\qquad\qquad\qquad + i \left( \sin(\Omega \tau) - \Omega \tau \right) 
    \Bigg] 
    d\Omega 
  \Bigg\}
\end{aligned}
\end{equation}
The quantum response function includes a reorganization correction, whereas the classical response lacks this correction due to the absence of dissipation. Moreover, the derivative term in the quantum linear response contains a zero-point energy contribution instead of thermal energy; that is, both quantum and classical responses become equivalent in the limit $\hbar \rightarrow 0$ when $\frac{1}{2}\hbar\omega_0 = k_BT$. Analytical expressions for the argument of the exponential of quantum linear response can be found elsewhere. \cite{jang2002temperature}

\subsection{Quantum non-linear response}
Perturbative treatment of light matter interaction allows us to write quantum third-order response as a function of time delays ($\tau_n$) can be written as follows\cite{Mukamel1995}

\begin{widetext}
\begin{equation}
\bm{R}_{\alpha_4\alpha_3\alpha_2\alpha_1}^{(3)}(\tau_3,\tau_2,\tau_1) = \Theta(\tau_3) \Theta(\tau_2) \Theta(\tau_1) \left( \frac{i}{\hbar} \right)^{3} \left\langle  \left[ {\bm{\mu}}_{\alpha_4}(\tau_1 + \tau_2 + \tau_3), \left[ {\bm{\mu}}_{\alpha_3}(\tau_1 + \tau_2), \left[ {\bm{\mu}}_{\alpha_2}(\tau_1), \left[ {\bm{\mu}}_{\alpha_1}(0), \hat{\rho}_\text{eq} \right] \right] \right] \right]  \right\rangle
\label{eq:Quantum_third_order}
\end{equation}
\end{widetext}
where, the time dependent dipole operator can be written as
\begin{equation}
\bm{\mu}\left(t\right)=\exp \left( \frac{i \hat{H}_{tot} t}{\hbar} \right) \bm{\mu} \exp \left( -\frac{i \hat{H}_{tot} t}{\hbar} \right)
\end{equation}
Now, using a linked diagram or time ordered cumulant expansion\cite{page1981separation,tonks1988diagrammatic,cho2001nonlinear,sung2001four,dutta2025nonlinear} we can obtain the non-rephasing and rephasing responses for harmonic system. The detailed derivation is provided in the Appendix C. The nonlinear resonances for anharmonic system can be obtained by expanding anharmonic transition frequency and transition dipole moments (unchanged for the first order)\cite{reppert2021diagrammatic} to the first order. Here we note only that in the limit $\hbar \rightarrow 0$, we obtain exactly the same expression as the classical nonlinear response, which validates our classical approach.

\section{Classical Linear absorption spectra}
\begin{figure}[h]
 \centering
 \includegraphics[height=6.3cm]{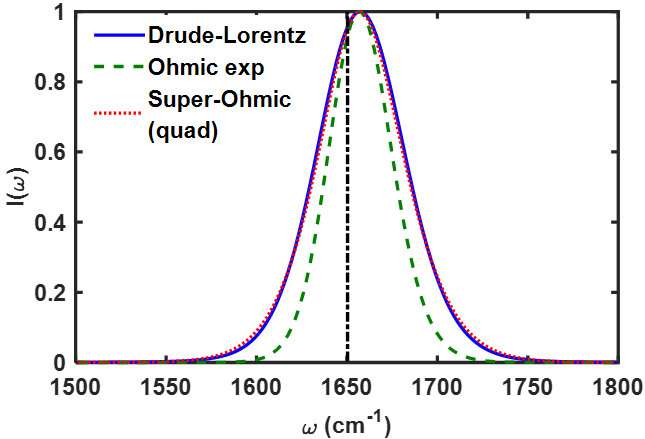}
\caption{Linear absorption spectra of an anharmonic oscillator for different bath spectral density with reorganization energy 2  {cm}$^{-1}$, cut-off frequency 10 {cm}$^{-1}$ and anharmonic constant $\Delta$=16 cm$^{-1}$. Line shapes are normalized to unity with respect to the individual maxima.}
 \label{fig:fig2}
\end{figure}

In this section, we evaluate and plot linear absorption spectra of an anharmonic oscillator for different bath spectral density.
Linear absorption spectra can be obtained from the Fourier transform of of linear response as follows
\begin{equation}
 I(\omega) = Im\Bigg[\int_0^{\infty} d\tau
e^{ i \omega \tau} R^{(1)}_{\alpha_1 \alpha_2}(\tau)\Bigg] 
\end{equation}
Analytical expressions for the linear response function show that the response is stable and exhibits no divergence, except in the case of the cubic super-Ohmic spectral density. Consequently, we plot the line shapes for Drude–Lorentz, Ohmic with exponential cutoff, and super-Ohmic (quadratic) spectral densities (Fig.~\ref{fig:fig2}), using a excitation frequency of 1650 cm$^{-1}$, reorganization energy of 2 cm$^{-1}$, cut-off frequency 10 cm$^{-1}$ and anharmonic constant of 16 cm$^{-1}$ relevant for amide I spectroscopy.
Let us first examine the features of the linear spectra (Fig.~\ref{fig:fig2}). Interestingly, the response functions for a super-Ohmic (cubic) spectral density cannot remove the linear divergence introduced (to first order in $\Delta$) by the anharmonicity. As a result, the plots of the linear spectra only include Ohmic (Lorentzian with exponential cutoff) and super-Ohmic (quadratic) spectral densities. In the case of the Drude–Lorentz spectral density, the bath oscillators contain both low- and high-frequency modes. The low-frequency modes govern the long-time dynamics and contribute to dephasing at long times, whereas the high-frequency modes are responsible for short-time dephasing. For an Ohmic spectral density with an exponential cut-off, the high-frequency bath modes are strongly suppressed. The absence of these high-frequency modes leads to a slower decay of the bath correlation function, and the resulting line shape is narrower compared to the Drude–Lorentz case. In contrast, the super-Ohmic (quadratic) spectral density maintains a balance between low- and high-frequency modes, producing a line shape similar to that of the Drude–Lorentz case. However, for the super-Ohmic (cubic) spectral density, the low-frequency modes are essentially absent. As a result, the linear divergence caused by anharmonicity cannot be lifted since there is no long-time dephasing. 

In addition, all line shapes manifest a frequency shift due to the derivative term in the response function arising from the first-order perturbation. To identify the origin of this shift, we note that, to lowest order in $\Delta$, the linear response function takes the form (see Eq.~\eqref{eq:linear_response})
\begin{equation}
\begin{aligned}
R^{1}(\tau) =\ & \left[ 
\frac{\sin(\omega_0 \tau)}{\omega_0} 
+ \frac{4 \Delta \tau k_B T \cos(\omega_0 \tau)}{\omega_0^2} 
\right] f(\tau),
\label{eq:linear_response_generic}
\end{aligned}
\end{equation}
where $f(\tau)$ is a non-negative function that depends on the bath spectral density. While the exact position of the absorption maximum depends on $f(\tau)$, a reasonable estimate is
\begin{align}
    \label{eq:tau_cross}
    \omega_\text{peak} = \omega_0 + \delta \omega = \frac{\pi}{ \tau_\text{cross}},
\end{align}
where $\tau_\text{cross}$ is the first zero-crossing time of the response function, a proxy for the dominant Fourier frequency. Since $f(\tau)$ is nonnegative, the zero-crossing time is simply the smallest positive time for which 
\begin{align}
    \frac{\sin(\omega_0 \tau_\text{cross})}{\omega_0} 
+ \frac{4 \Delta \tau_\text{cross} k_B T \cos(\omega_0 \tau_\text{cross})}{\omega_0^2} = 0 .
\label{eq:tau_cross_condition}
\end{align}
Anticipating that the frequency shift will be small relative to $\omega_0$, we solve Eq.~\eqref{eq:tau_cross} to first order in $\frac{\delta \omega}{\omega_0}$ as 
\begin{align}
    \tau_\text{cross} = \frac{\pi}{\omega_0 + \delta \omega } \approx \frac{\pi}{\omega_0} \left( 1 - \frac{\delta \omega}{\omega_0} \right)
\end{align}
to obtain from Eq.~\eqref{eq:tau_cross_condition}
\begin{align}
    \tan \left (\pi \left( 1 - \frac{\delta \omega}{\omega_0} \right) \right ) =- \frac{4 \Delta k_B T}{\omega_0} \frac{\pi}{\omega_0} \left( 1 - \frac{\delta \omega}{\omega_0} \right) .
\end{align}
To lowest order in $\frac{\delta \omega}{\omega_0}$ and in $\Delta$, this is satisfied when
\begin{align}
    \label{eq:shift_approx}
    \delta \omega = \frac{4 \Delta k_B T}{\omega_0} .
\end{align}
Using the parameter sets of Fig.~\ref{fig:fig2}, the calculated shift is about $7~\text{cm}^{-1}$, while Eq.~\eqref{eq:shift_approx} estimates approximately $8~\text{cm}^{-1}$, indicating that our approximations are reasonable for the present system. Note that the predicted peak shift is independent of the strength of the system-bath coupling and should thus not be thought of as a reorganization energy, which does not exist in our classical model.\cite{PhysRevA.111.022210} Instead, the shift is proportional to the temperature and the anharmonicity and may be understood physically as a result of the fact that thermal excitation allows the system to explore anharmonic regions of the potential that would be inaccessible at $T = 0$.\cite{10.1063/1.5017985} 

\section{Classical two dimensional spectra}
In this section, we present 2D IR spectra obtained from the classical approach for different bath spectral densities using parameters relevant to Amide I systems. To obtain the 2D correlation plot, we need to Fourier transform the non-rephasing and rephasing response as follows 
\begin{equation}
\begin{split}
 S^{\text{NR/R}}(\omega_1, \tau_2, \omega_3) = 
& \int_0^{\infty} d\tau_1 \int_0^{\infty} d\tau_3 \,
e^{\mp i \omega_1 \tau_1} e^{i \omega_3 \tau_3} \\
& \times R_{\alpha_1 \alpha_2 \alpha_3 \alpha_4}^{\text{(NR/R)}}(\tau_3, \tau_2, \tau_1)   
\end{split}
\end{equation}
where the “-” represents to nonrephasing and the “+” to rephasing.
2D correlation surface can be written as
\begin{equation}
S^{\text{corr}} = \text{Im} \Bigg[ S^{\text{NR}} + S^{\text{R}} \Bigg]
\end{equation}
\begin{figure*}
    \centering
    \includegraphics[height=9cm]{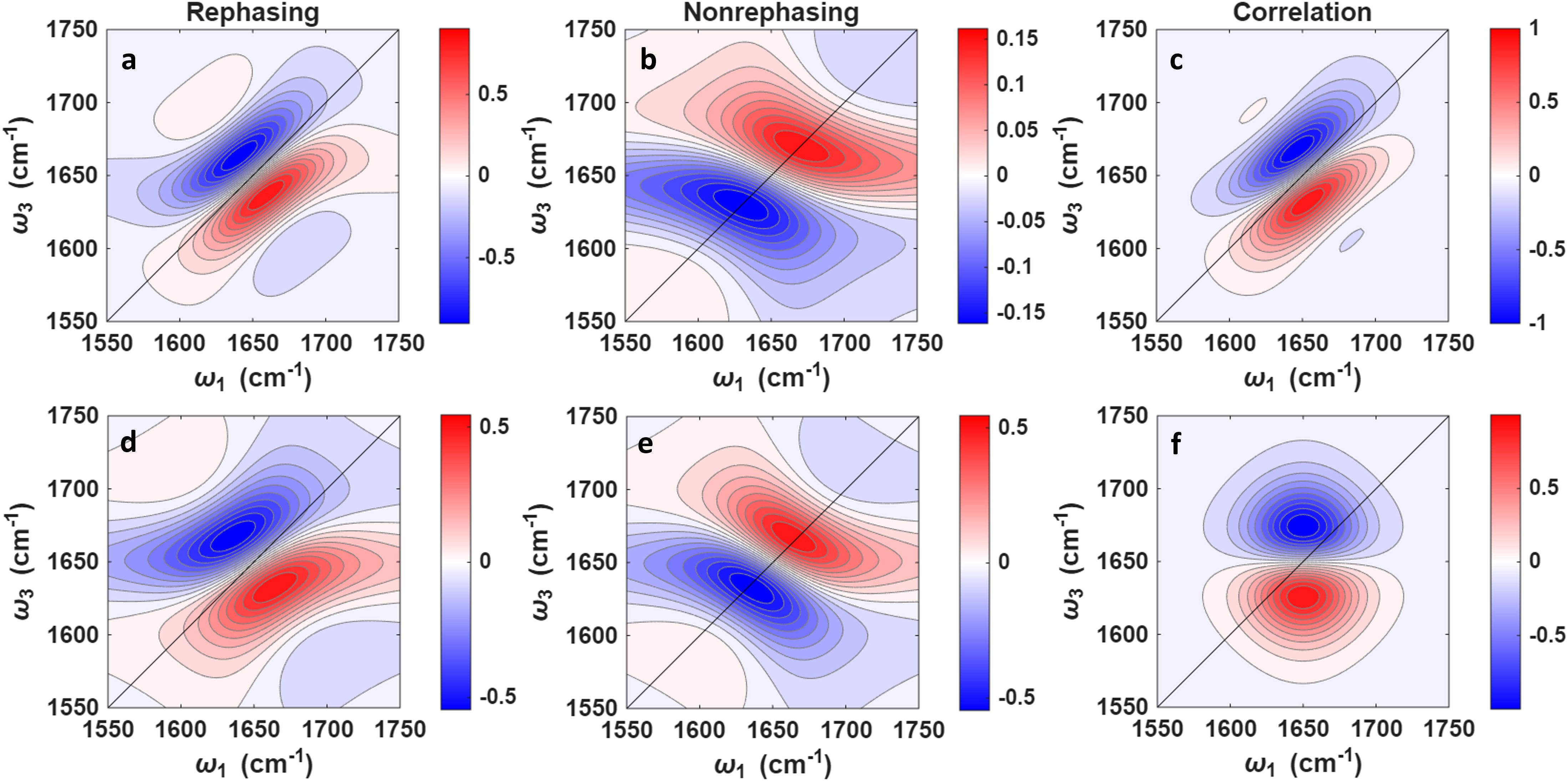}
    \caption{2D IR spectra for Drude-Lorentz bath spectral density with either $\tau_2=0$ (Frames a - c) or $\tau_2=10$ ps (Frames d - f). The other parameters are same as that of absorption spectra. Frames (a) and (d) are rephasing surfaces; Frames (b) and (e) are nonrephasing; and Frames (c) and (f) are correlation surfaces.}
    \label{fig:2dir_drude_lorentz}
\end{figure*}
\begin{figure*}
    \centering
    \includegraphics[height=9cm]{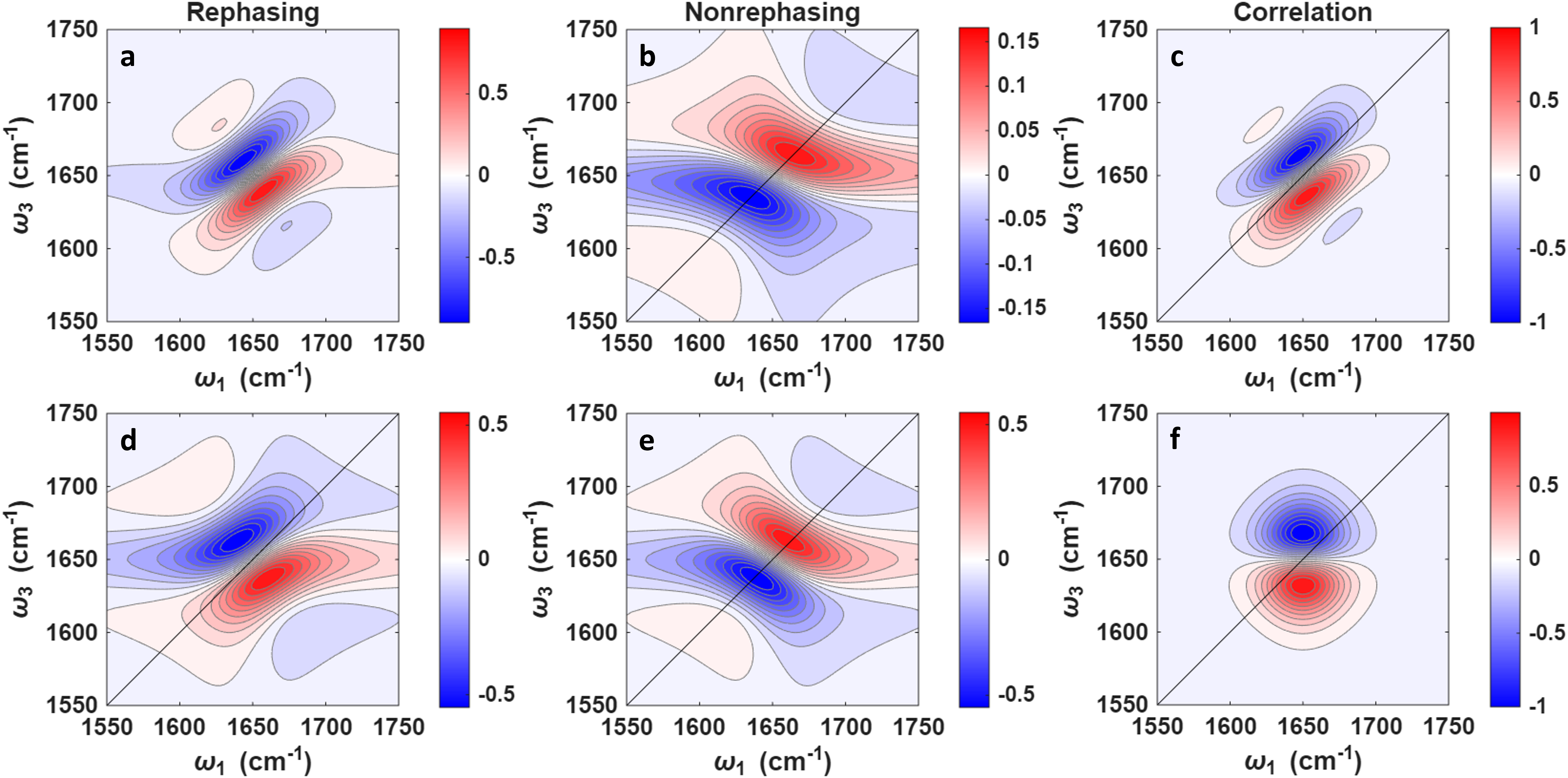}
    \caption{2D IR spectra for Ohmic bath spectral density with either $\tau_2=0$ (Frames a - c) or $\tau_2=10$ ps (Frames d - f). The other parameters are same as that of absorption spectra. Frames (a) and (d) are rephasing surfaces; Frames (b) and (e) are nonrephasing; and Frames (c) and (f) are correlation surfaces.}
    \label{fig:2dir_ohmic}
\end{figure*}
\begin{figure*}
    \centering
    \includegraphics[height=9cm]{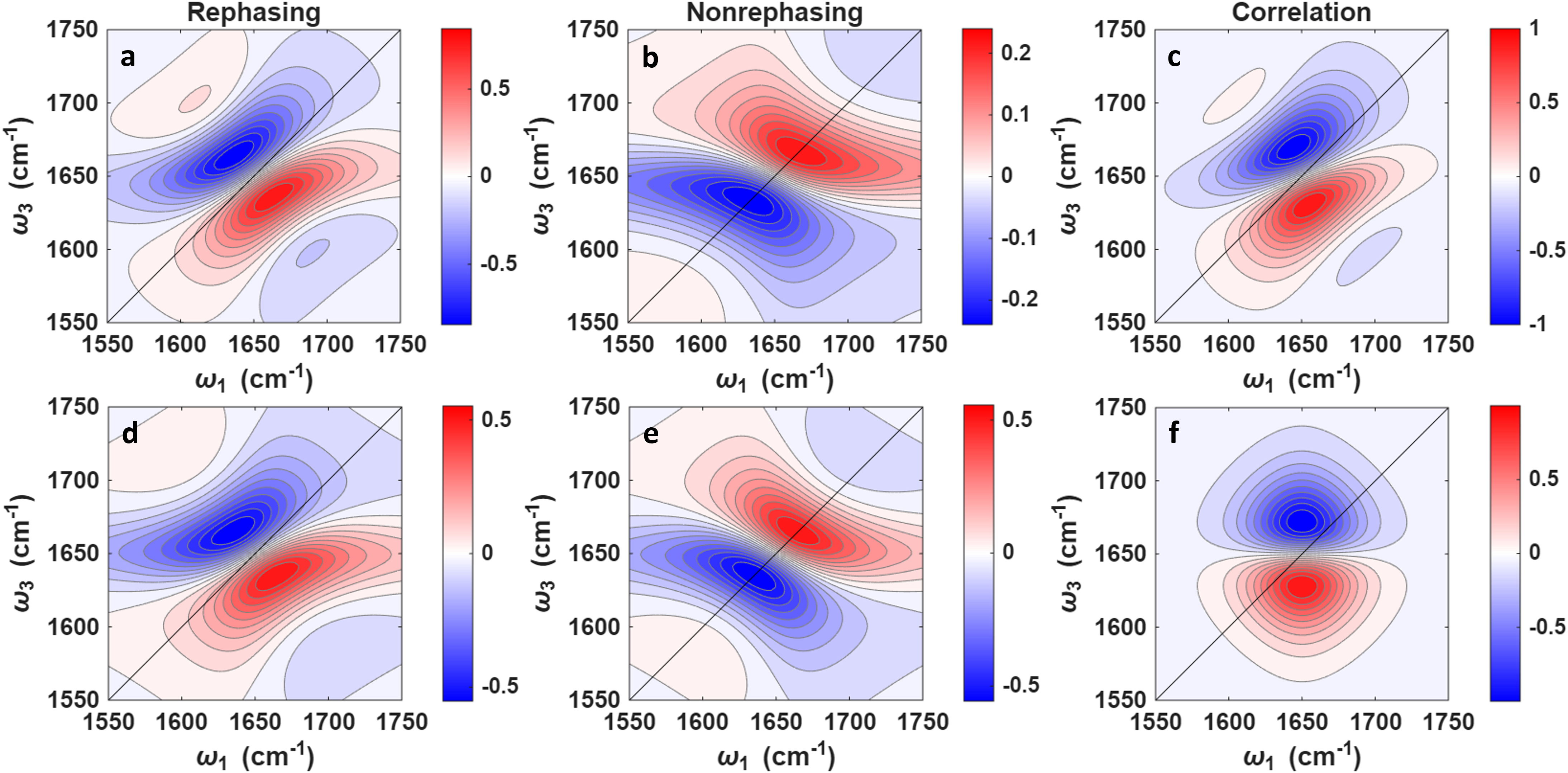}
    \caption{2D IR spectra for super-Ohmic (quadratic) bath spectral density with either $\tau_2=0$ (Frames a - c) or $\tau_2=10$ ps (Frames d - f). The other parameters are same as that of absorption spectra. Frames (a) and (d) are rephasing surfaces; Frames (b) and (e) are nonrephasing; and Frames (c) and (f) are correlation surfaces.}
    \label{fig:2dir_super_ohmic}
\end{figure*}

In the earlier section, we showed analytically that the presence of a classical bath stabilizes the classical nonlinear response function. In the present section, to illustrate the spectroscopic manifestations of different bath spectral densities, Figs.~\ref{fig:2dir_drude_lorentz}, \ref{fig:2dir_ohmic}, and \ref{fig:2dir_super_ohmic} present the frequency-domain rephasing and nonrephasing contributions, along with the corresponding 2D correlation plots. Let us first examine the general dynamical features of all the 2D correlation plots. At short population times ($\tau_2 \approx 0$), the 2D IR spectra exhibit elongation along the diagonal because the bath-induced frequency fluctuations are still small. The system has not yet experienced significant dephasing or environmental modulation. The bath-induced fluctuations are still small on this timescale, so the environment effectively appears static, so the excitation and detection frequencies remain effectively correlated, producing diagonally stretched peaks. In this regime, the rephasing contribution is stronger than the non-rephasing contribution because it can partially refocus the phase evolution accumulated during the coherence period, leading to constructive interference along the diagonal. As $\tau_2$ increases, the dynamics of the bath induces larger frequency fluctuations, gradually modulating the system and diminishing phase memory. Consequently, the rephasing and nonrephasing contributions become comparable at long $\tau_2$, and the peaks evolve toward a more circular shape, reflecting homogeneous broadening arising from the system-bath interactions. In other words, the decay of diagonal elongation into a more circular peak shape with increasing $\tau_2$ is a direct signature of spectral diffusion, arising from the progressive loss of frequency correlation.

Similar to the absorption spectra shown in Fig.~\ref{fig:fig2}, the 2D correlation plots manifest a comparable feature: the lobes are narrowed for the Ohmic spectral density with an exponential cutoff, which can be attributed to the absence of significant high-frequency bath modes. This narrowing is observed for both the red lobes (associated with ground-state bleaching and stimulated emission) and the blue lobes (corresponding to excited-state absorption) at both short and long population times. Although the 1D absorption spectra for the Drude–Lorentz and super-Ohmic (quadratic) spectral densities are nearly indistinguishable, their 2D correlation plots reveal broader lobes for the super-Ohmic case.

This difference arises because linear absorption spectra reflect only the overall line shape, effectively averaging over bath-induced fluctuations. Although the Drude-Lorentz spectral density has a long tail that extends to high frequencies, the weight of these high-frequency modes is small, so their effect on the absorption spectrum is minimal. In contrast, the super-Ohmic spectral density has stronger high-frequency modes, which induce faster dephasing and larger frequency fluctuations. These fluctuations affect the correlations between excitation and detection frequencies and the population-time dynamics, which are directly probed in 2D spectroscopy. Consequently, the differences between Drude–Lorentz and super-Ohmic bath spectral density, subtle in linear absorption, become pronounced in 2D spectra as a broadening of the lobes along both axes.
\section{Conclusion}
The key findings of this work can be summarized as follows. We first derived the linear response under the weak-anharmonicity approximation for a classical anharmonic oscillator and showed that system–bath interactions can remove the fundamental linear divergence introduced by anharmonicity. Our analytical results further demonstrate that, depending on whether the bath is fast or slow, one can recover the homogeneous or inhomogeneous broadening limit for an Ohmic spectral density. Importantly, the form of the bath spectral density plays a crucial role in stabilizing the response, as the analysis highlights the significance of low-frequency modes. In particular, the absence of low-frequency modes in a super-Ohmic (cubic) spectral density fails to eliminate the divergence, whereas the absence of high-frequency modes in an Ohmic spectral density with an exponential cutoff produces a narrowed absorption spectrum.

The quantum linear response yields the same response function as the classical case at first-order anharmonicity in the limit $\hbar\rightarrow0$. Moreover, first-order perturbative effects manifest as shifts in the linear spectra, offering further insight into the role of anharmonicity and bath characteristics.

We then extended the diagrammatic approach to obtain the nonlinear response for a classical anharmonic oscillator coupled to a bath, building on earlier work for isolated systems~\cite{reppert2021diagrammatic,reppert2023equivalence}. Similar to the case of classical linear response, the analytical expression for the nonlinear response shows equivalence with the quantum response in the limit $\hbar \rightarrow 0$. The derived expressions reveal that coupling to a bath can remove the linear $\tau_3$ divergence, with the outcome strongly dependent on the form of the bath spectral density, analogous to the linear-response case. For an Ohmic spectral density, our analysis shows that in the static-disorder or slow-bath limit, the non-rephasing response remains stable, whereas the rephasing response exhibits a divergence along the $\tau_1 = \tau_3$ axis.  

Two-dimensional correlation plots for parameters relevant to the Amide I band further demonstrate that our fully classical framework captures $\tau_2$-dependent spectral diffusion consistent with quantum treatments: at short $\tau_2$, the spectra are elongated along the diagonal, similar to static-disorder effects, while at long $\tau_2$, the lobes evolve toward a circular shape, characteristic of homogeneous broadening. Unlike the linear spectra, the nonlinear 2D spectra exhibit more pronounced differences in broadening depending on the bath spectral density, thereby highlighting the sensitivity of higher-order response to environmental characteristics.  

In closing, this work resolves the fundamental question of linear divergence in the classical nonlinear response by identifying its solution in the system–bath interaction, with the behavior depending critically on the nature of the bath spectral density. Although our analysis is based on the weak-anharmonicity approximation and single-mode models, it remains highly relevant to 2D IR spectroscopy of realistic systems and is particularly suited for cases where anharmonicity can be treated perturbatively, highlighting a new direction and the broader applicability of classical approaches. It is worth noting that the classical description does not capture the reorganization shift, a feature intrinsic to quantum mechanics, reflecting a fundamental distinction rather than a limitation of the method.\cite{PhysRevA.111.022210}

\begin{acknowledgments}
This research was supported by the National Science Foundation under Grant CHE-2236625.
\end{acknowledgments}

\section*{Data Availability Statement}

The data that support the findings of this study are available within the article.

\appendix
\section{Simplified classical propagator}
We define a new Liouvillian
\begin{equation}
{{\tilde{L}}_{SB}}= -i\sum\limits_{k}{\frac{{{\alpha }_{k}}}{\Omega _{k}^{2}}{{P}_{k}}\frac{\partial }{\partial \theta }}
\end{equation}

 The new Liouvillian has following commutation with bath Liouvillian
\begin{equation}
\begin{split}
  & \left[ {{{\tilde{L}}}_{SB}},{{{\hat{L}}}_{B}} \right]=-i{{{\hat{L}}}_{SB}} \\ 
 & \left[ {{{\tilde{L}}}_{SB}},\left[ {{{\tilde{L}}}_{SB}},{{{\hat{L}}}_{B}} \right] \right]=0 \\ 
 & \left[ {{{\tilde{L}}}_{SB}},\left[ {{{\tilde{L}}}_{SB}},\left[ {{{\tilde{L}}}_{SB}},{{{\hat{L}}}_{B}} \right] \right] \right]=0  
\end{split}
\end{equation}

Now, using the commutations above and insert this into Baker-Campbell-Haudsroff formula, we obtain the following propagator
\begin{equation}
  {{e}^{-i\left( \hat{L}_{SB} + \hat{L}_{B} \right) \tau}} 
  = {{e}^{-i \left( e^{\tilde{L}_{SB}} \hat{L}_{B} e^{-\tilde{L}_{SB}} \right) \tau}} 
  = e^{i \tilde{L}_{SB}} e^{-i \hat{L}_{B} \tau} e^{-i \tilde{L}_{SB}} 
\end{equation}

The time dependence of the new system-bath Liouvillian can be expressed in the following interaction representation as
\begin{equation}
\begin{split}
 & {{e}^{-i{{{\hat{L}}}_{B}}{{\tau }}}}{{{\tilde{L}}}_{SB}}{{e}^{i{{{\hat{L}}}_{B}}{{\tau }}}}={{{\tilde{L}}}_{SB}}\left( {{\tau }} \right) \\ 
 & {{e}^{-i{{{\hat{L}}}_{B}}{{\tau }}}}{{e}^{-i{{{\tilde{L}}}_{SB}}}}{{e}^{i{{{\hat{L}}}_{B}}{{\tau }}}}={{e}^{-i{{{\tilde{L}}}_{SB}}\left( {{\tau }} \right)}} 
\end{split}
\end{equation}
As a result, total propagator becomes 
\begin{equation}
e^{-i\left( \hat{L}_{S}^{(0)} + \hat{L}_{SB} + \hat{L}_{B} \right)\tau} = e^{-i \hat{L}_{S}^{(0)} \tau} \, e^{i \tilde{L}_{SB}} \, e^{-i \tilde{L}_{SB} \left(\tau\right)} \, e^{-i \hat{L}_{B} \tau}
\end{equation}
\section{Classical non-linear response for the generalized spectral density}
Using the generalized spectral density with exponential cut-off Eq.~\eqref{eq:gen_spec_den}, we obtain the following expressions of non-rephasing and rephasing responses for Ohmic ($n=1$) and super-Ohmic($n=2$ and $n=3$) spectral density as
for $n=1$,
\begin{widetext}
\begin{equation}
\begin{aligned}
R_{\alpha_1 \alpha_2 \alpha_3 \alpha_4}^{(NR)}(\tau_1, \tau_2, \tau_3) &= \frac{\mu_{\alpha_1}^{(1)} \mu_{\alpha_2}^{(1)} \mu_{\alpha_3}^{(1)} \mu_{\alpha_4}^{(1)} \Delta \tau_3}{\omega_0^2} 
\, e^{-i \omega_0 (\tau_1 + \tau_3)} \\
&\quad \exp\Big[-\frac{A_1}{\beta} \Big[ 
    \tau_1 \tan^{-1}(\Omega_c \tau_1) 
    + \tau_2 \tan^{-1}(\Omega_c \tau_2) 
    + \tau_3 \tan^{-1}(\Omega_c \tau_3) \\
&\quad 
    - (\tau_1 + \tau_2) \tan^{-1}(\Omega_c (\tau_1 + \tau_2)) 
    - (\tau_2 + \tau_3) \tan^{-1}(\Omega_c (\tau_2 + \tau_3))  \\
&\quad 
    + (\tau_1 + \tau_2 + \tau_3) \tan^{-1}(\Omega_c (\tau_1 + \tau_2 + \tau_3)) 
\Big]\Big]\\
&\quad \times \bigg[{\frac{(1 + \Omega_c^2\tau_1^2)(1 + \Omega_c^2\tau_2^2)(1 + \Omega_c^2\tau_3^2)(1 + \Omega_c^2(\tau_1 + \tau_2 + \tau_3)^2)}{(1 + \Omega_c^2(\tau_1 + \tau_2)^2)(1 + \Omega_c^2(\tau_2 + \tau_3)^2)}}\bigg]^{\frac{A_1}{2\beta\Omega_c}}+\text{c.c.}
\end{aligned}
\end{equation}
\end{widetext}
\begin{widetext}
\begin{equation}
\begin{aligned}
R_{\alpha_1 \alpha_2 \alpha_3 \alpha_4}^{(R)}(\tau_1, \tau_2, \tau_3) &= \frac{\mu_{\alpha_1}^{(1)} \mu_{\alpha_2}^{(1)} \mu_{\alpha_3}^{(1)} \mu_{\alpha_4}^{(1)} \Delta \tau_3}{\omega_0^2} 
\, e^{-i \omega_0 (\tau_3 - \tau_1)} \\
& \quad \exp\Big[-\frac{A_1}{\beta} \Big[ 
    \tau_1 \tan^{-1}(\Omega_c \tau_1) 
    - \tau_2 \tan^{-1}(\Omega_c \tau_2) 
    + \tau_3 \tan^{-1}(\Omega_c \tau_3) \\
&\quad 
    + (\tau_1 + \tau_2) \tan^{-1}(\Omega_c (\tau_1 + \tau_2)) 
    + (\tau_2 + \tau_3) \tan^{-1}(\Omega_c (\tau_2 + \tau_3))  \\
&\quad 
    - (\tau_1 + \tau_2 + \tau_3) \tan^{-1}(\Omega_c (\tau_1 + \tau_2 + \tau_3)) 
\Big]\Big]  \\
&\quad \times \bigg[{\frac{(1 + \Omega_c^2\tau_1^2)(1 + \Omega_c^2\tau_3^2)(1 + \Omega_c^2(\tau_1 + \tau_2)^2)(1 + \Omega_c^2(\tau_2 + \tau_3)^2)}{(1 + \Omega_c^2\tau_2^2) (1 + \Omega_c^2(\tau_1 + \tau_2 + \tau_3)^2)}}\bigg]^{\frac{A_1}{2\beta\Omega_c}}+\text{c.c.}
\end{aligned}
\end{equation}

for n=2,

\begin{align}
R_{\alpha_1 \alpha_2 \alpha_3 \alpha_4}^{(NR)}(\tau_1, \tau_2, \tau_3) &= \frac{\mu_{\alpha_1}^{(1)} \mu_{\alpha_2}^{(1)} \mu_{\alpha_3}^{(1)} \mu_{\alpha_4}^{(1)} \Delta \tau_3}{\omega_0^2} 
\, e^{-i \omega_0 (\tau_1 + \tau_3)} \notag \\
&\quad \times \left[\frac{(1 + \Omega_c^2\tau_1^2)(1 + \Omega_c^2\tau_2^2)(1 + \Omega_c^2\tau_3^2)(1 + \Omega_c^2(\tau_1 + \tau_2 + \tau_3)^2)}{(1 + \Omega_c^2(\tau_1 + \tau_2)^2)(1 + \Omega_c^2(\tau_2 + \tau_3)^2)}\right]^{-\frac{A_2}{2\beta\Omega_c}} + \text{c.c.}
\end{align}
\end{widetext}

\begin{widetext}
\begin{equation}
\begin{aligned}
R_{\alpha_1 \alpha_2 \alpha_3 \alpha_4}^{(R)}(\tau_1, \tau_2, \tau_3) &= \frac{\mu_{\alpha_1}^{(1)} \mu_{\alpha_2}^{(1)} \mu_{\alpha_3}^{(1)} \mu_{\alpha_4}^{(1)} \Delta \tau_3}{\omega_0^2} 
\, e^{-i \omega_0 (\tau_3 - \tau_1)} \\
&\quad \times \bigg[{\frac{(1 + \Omega_c^2\tau_1^2)(1 + \Omega_c^2\tau_3^2)(1 + \Omega_c^2(\tau_1 + \tau_2)^2)(1 + \Omega_c^2(\tau_2 + \tau_3)^2)}{(1 + \Omega_c^2\tau_2^2) (1 + \Omega_c^2(\tau_1 + \tau_2 + \tau_3)^2)}}\bigg]^{-\frac{A_2}{2\beta\Omega_c}}+\text{c.c.}
\end{aligned}
\end{equation}

for n=3,

\begin{equation}
\begin{aligned}
R_{\alpha_1 \alpha_2 \alpha_3 \alpha_4}^{(NR)}(\tau_1, \tau_2, \tau_3) &= \frac{\mu_{\alpha_1}^{(1)} \mu_{\alpha_2}^{(1)} \mu_{\alpha_3}^{(1)} \mu_{\alpha_4}^{(1)} \Delta \tau_3}{\omega_0^2} 
\, e^{-i \omega_0 (\tau_1 + \tau_3)} \\
&\quad \exp\Big[-\frac{A_3}{3\beta\Omega_c^2} \Big[\frac{\Omega_c^3\tau_1^2}{1+\Omega_c^2\tau_1^2} 
+\frac{\Omega_c^3\tau_2^2}{1+\Omega_c^2\tau_2^2}
+\frac{\Omega_c^3\tau_3^2}{1+\Omega_c^2\tau_3^2} \\
    &\quad - \frac{\Omega_c^3\left(\tau_1+\tau_2\right)^2}{1+\Omega_c^2\left(\tau_1+\tau_2\right)^2}
    -\frac{\Omega_c^3\left(\tau_2+\tau_3\right)^2}{1+\Omega_c^2\left(\tau_2+\tau_3\right)^2}
    + \frac{\Omega_c^3\left(\tau_1+\tau_2+\tau_3\right)^2}{1+\Omega_c^2\left(\tau_1+\tau_2+\tau_3\right)^2} \Big]\Big]+\text{c.c.}
\end{aligned}
\end{equation}
\end{widetext}

\begin{widetext}
\begin{equation}
\begin{aligned}
R_{\alpha_1 \alpha_2 \alpha_3 \alpha_4}^{(R)}(\tau_1, \tau_2, \tau_3) &= \frac{\mu_{\alpha_1}^{(1)} \mu_{\alpha_2}^{(1)} \mu_{\alpha_3}^{(1)} \mu_{\alpha_4}^{(1)} \Delta \tau_3}{\omega_0^2} 
\, e^{-i \omega_0 (\tau_3 - \tau_1)} \\
&\quad \times \exp\Big[-\frac{A_3}{3\beta\Omega_c^2} \Big[\frac{\Omega_c^3\tau_1^2}{1+\Omega_c^2\tau_1^2} 
-\frac{\Omega_c^3\tau_2^2}{1+\Omega_c^2\tau_2^2}
+\frac{\Omega_c^3\tau_3^2}{1+\Omega_c^2\tau_3^2}  \\
    &\quad + \frac{\Omega_c^3\left(\tau_1+\tau_2\right)^2}{1+\Omega_c^2\left(\tau_1+\tau_2\right)^2}
+\frac{\Omega_c^3\left(\tau_2+\tau_3\right)^2}{1+\Omega_c^2\left(\tau_2+\tau_3\right)^2}
    - \frac{\Omega_c^3\left(\tau_1+\tau_2+\tau_3\right)^2}{1+\Omega_c^2\left(\tau_1+\tau_2+\tau_3\right)^2} \Big]\Big]+\text{c.c.}
\end{aligned}
\end{equation}
\end{widetext}

\section{Derivation of quantum non-linear response function}

Repeated insertion of identity operator $\sum_a |a\rangle \langle a|$ ($a$ = 0, 1 and 2) on the left of dipole operators and by using diagrammatic approach for 3 level system, we obtain the non-rephasing contribution as
\begin{widetext}
\begin{equation}
\begin{split}
\bm{R}_{NR,\alpha_4\alpha_3\alpha_2\alpha_1}^{(3)}(\tau_3,\tau_2,\tau_1) = & \; \left(\frac {i}{\hbar}\right)^3 \Bigg[ {{\mu}}_{\alpha_1}^{10} {{\mu}}_{\alpha_2}^{01} {{\mu}}_{\alpha_3}^{10} {{\mu}}_{\alpha_4}^{01} \; \text{Tr}_\text{B} \Bigg[ \rho_0^{00} \exp \left( \frac{i H_0 (\tau_3 + \tau_2)}{\hbar} \right)   \\
& \exp\left( -\frac{i H_1 (\tau_3 + \tau_2)}{\hbar} \right) \exp \left( \frac{i H_1 \tau_2}{\hbar} \right) \exp \left( -\frac{i H_0 \tau_2}{\hbar} \right) \\
& \exp \left(- \frac{i H_1 \tau_1}{\hbar} \right) \exp \left(\frac{i H_0 \tau_1}{\hbar} \right) \Bigg] \\
& + {{\mu}}_{\alpha_1}^{10} {{\mu}}_{\alpha_2}^{01} {{\mu}}_{\alpha_3}^{10} {{\mu}}_{\alpha_4}^{01} \; \text{Tr}_\text{B} \Bigg[ \rho_0^{00} \exp \left( \frac{i H_0 \tau_3}{\hbar} \right)\exp\left( -\frac{i H_1 \tau_3}{\hbar} \right)  \\
& \exp\left( -\frac{i H_1 (\tau_1 + \tau_2)}{\hbar} \right) \exp \left( \frac{i H_0 (\tau_1 + \tau_2)}{\hbar} \right)  \\
& \exp \left( -\frac{i H_0 \tau_2}{\hbar} \right) \exp \left( \frac{i H_1 \tau_2}{\hbar} \right) \Bigg] \\
& - {{\mu}}_{\alpha_1}^{10} {{\mu}}_{\alpha_2}^{01} {{\mu}}_{\alpha_3}^{21} {{\mu}}_{\alpha_4}^{12} \; \text{Tr}_\text{B} \Bigg[ \rho_0^{00} \exp \left( \frac{i H_0 \tau_3}{\hbar} \right)\exp \left(- \frac{i H_2 \tau_3}{\hbar} \right)  \\
& \exp \left( -\frac{i H_1 (\tau_1 + \tau_2)}{\hbar} \right) \exp \left( \frac{i H_0 (\tau_1 + \tau_2)}{\hbar} \right) \\
& \exp \left( -\frac{i H_0 \tau_2}{\hbar} \right) \exp \left( \frac{i H_1 \tau_2}{\hbar} \right) \exp \left( \frac{i H_1 \tau_3}{\hbar} \right) \exp \left(- \frac{i H_0 \tau_3}{\hbar} \right)  -c.c.\Bigg] \Bigg]
\end{split}
\end{equation}
\end{widetext}
One can rewrite the following expression as follows
\begin{widetext}
\begin{equation}
\begin{split}
\bm{R}_{NR,\alpha_4\alpha_3\alpha_2\alpha_1}^{(3)}(\tau_3,\tau_2,\tau_1) 
= & \; \left(\frac {i}{\hbar}\right)^3 \Bigg\{ 
{{\mu}}_{\alpha_1}^{10} {{\mu}}_{\alpha_2}^{01} {{\mu}}_{\alpha_3}^{10} {{\mu}}_{\alpha_4}^{01} 
\exp\left[-i\omega_{10}(\tau_3+\tau_1)\right] \\
& \times \Bigg[ \text{Tr}_\text{B} \Bigg( 
\exp_{+} \left[ -\frac{i}{\hbar} \int_0^{\tau_3 + \tau_2 } ds_0 \delta {H}_{10}(s_0) \right]  
\exp_{-} \left[ \frac{i}{\hbar} \int_0^{\tau_2} ds_1 \delta {H}_{10}(s_1) \right]  \\
& \quad \times \exp_{-} \left[- \frac{i}{\hbar} \int_0^{\tau_1} ds_2 \delta {H}_{10}(-s_2) \right]
\Bigg) \\
& +\text{Tr}_\text{B} \Bigg( 
\exp_{+} \left[ -\frac{i}{\hbar} \int_0^{\tau_3} ds_0 \delta {H}_{10}(s_0) \right]  
\exp_{-} \left[- \frac{i}{\hbar} \int_0^{\tau_1 + \tau_2} ds_1 \delta {H}_{10}(-s_1) \right] \\
& \quad \times \exp_{+} \left[- \frac{i}{\hbar} \int_0^{\tau_2} ds_2 \delta {H}_{10}(-s_2) \right]
\Bigg)\Bigg] \\
& - {{\mu}}_{\alpha_1}^{10} {{\mu}}_{\alpha_2}^{01} {{\mu}}_{\alpha_3}^{21} {{\mu}}_{\alpha_4}^{12} 
\exp\left[-i\omega_{10}\tau_1\right] \exp\left[-i\omega_{21}\tau_3\right]  \\
& \times \text{Tr}_\text{B} \Bigg[  
\exp_{+} \left[- \frac{i}{\hbar} \int_0^{\tau_3} ds_0 \delta {H}_{20}(s_0) \right]
\exp_{-} \left[ -\frac{i}{\hbar} \int_0^{\tau_2 + \tau_1} ds_0 \delta {H}_{10}(-s_1) \right] \\
& \quad \times   
\exp_{+} \left[ \frac{i}{\hbar} \int_0^{\tau_2} ds_1 \delta {H}_{10}(-s_2) \right] \exp_{-} \left[ \frac{i}{\hbar} \int_0^{\tau_3} ds_3 \delta {H}_{10}(s_3) \right] \Bigg] -c.c.\Bigg\}
\end{split}
\end{equation}

Similarly, the rephasing contribution can be written as

\begin{equation}
\begin{split}
\bm{R}_{R,\alpha_4\alpha_3\alpha_2\alpha_1}^{(3)}(\tau_3,\tau_2,\tau_1) = & \; \left(\frac {i}{\hbar}\right)^3 \Bigg\{ 
{{\mu}}_{\alpha_1}^{10} {{\mu}}_{\alpha_2}^{01} {{\mu}}_{\alpha_3}^{10} {{\mu}}_{\alpha_4}^{01} 
\exp\left[-i\omega_{10}(\tau_3-\tau_1)\right] \\
& \times \Bigg[ \text{Tr}_\text{B} \Bigg( 
\exp_{+} \left[ -\frac{i}{\hbar} \int_0^{\tau_3} ds_0 \delta {H}_{10}(s_0) \right]  
\exp_{+} \left[ \frac{i}{\hbar} \int_0^{\tau_1+\tau_2} ds_1 \delta {H}_{10}(-s_1) \right]  \\
& \quad \times \exp_{-} \left[- \frac{i}{\hbar} \int_0^{\tau_2} ds_2 \delta {H}_{10}(-s_2) \right]
\Bigg) \\
& +\text{Tr}_\text{B} \Bigg( 
\exp_{+} \left[ -\frac{i}{\hbar} \int_0^{\tau_2+\tau_3} ds_0 \delta {H}_{10}(s_0) \right]  
\exp_{+} \left[ \frac{i}{\hbar} \int_0^{\tau_1} ds_1 \delta {H}_{10}(-s_1) \right] \\
& \quad \times \exp_{-} \left[ \frac{i}{\hbar} \int_0^{\tau_2} ds_2 \delta {H}_{10}(s_2) \right]
\Bigg)\Bigg] \\
& - {{\mu}}_{\alpha_1}^{10} {{\mu}}_{\alpha_2}^{01} {{\mu}}_{\alpha_3}^{21} {{\mu}}_{\alpha_4}^{12} 
\exp\left[i\omega_{10}\tau_1\right] \exp\left[-i\omega_{21}\tau_3\right] \\
& \times \text{Tr}_\text{B} \Bigg[ 
\exp_{+} \left[ \frac{i}{\hbar} \int_0^{\tau_3} ds_0 \delta {H}_{10}(-s_0) \right]
\exp_{-} \left[ \frac{i}{\hbar} \int_0^{\tau_2 + \tau_1} ds_0 \delta {H}_{10}(s_1) \right] \\
& \quad \times   
\exp_{+} \left[- \frac{i}{\hbar} \int_0^{\tau_2} ds_1 \delta {H}_{10}(s_2) \right] \exp_{-} \left[-\frac{i}{\hbar} \int_0^{\tau_3} ds_3 \delta {H}_{20}(-s_3) \right]\Bigg]  -c.c.\Bigg\}
\end{split}
\end{equation}
\end{widetext}
Here, the time-dependent fluctuating part of the Hamiltonian is defined as
\begin{equation}
\label{eq:interaction}
\begin{aligned}
  & {{e}^{i{{H}^0_{tot}}\tau /\hbar }}\left( {{{{H}}}^x_{tot}}-{{{{H}}}^0_{tot}} \right){{e}^{-i{{H}^0_{tot}}\tau /\hbar }} \\ 
 & =\hbar {\omega }_{x0} + \delta {{{{H}}}_{x0}}\left( \tau  \right) 
\end{aligned}
\end{equation}
and ``+" and ``-" subscripts of the exponential indicate positive and negative time ordering respectively.
Now, using linked diagram or cumulant expansion theorem we obtain the following expression for non-rephasing and rephasing response function. We also assumed $\delta_{20}=2\delta_{10}$.

\begin{widetext}
\begin{equation}
\begin{split}
    \bm{R}_{NR,\alpha_4\alpha_3\alpha_2\alpha_1}^{(3)}(\tau_3,\tau_2,\tau_1) 
    = & \; \left(\frac {i}{\hbar}\right)^3 
    \Bigg[ 2{{\mu}}_{\alpha_1}^{10} {{\mu}}_{\alpha_2}^{01} {{\mu}}_{\alpha_3}^{10} {{\mu}}_{\alpha_4}^{01} 
    \exp\left[-i\omega_{10}(\tau_3+\tau_1)\right] \\
    & \quad \times \exp \Bigg( -\int_0^{\infty} d\Omega \coth{\left(\frac{\beta\hbar\Omega}{2}\right)} 
    \frac{C''\left(\Omega\right)}{\Omega^2} \bigg(2-\cos\Omega\tau_1-\cos\Omega\tau_2-\cos\Omega\tau_3  \\
    & \qquad +\cos\Omega(\tau_1+\tau_2)+\cos\Omega(\tau_2+\tau_3)-\cos\Omega(\tau_1+\tau_2+\tau_3) \bigg)  \\
    & \quad +i\int_0^{\infty} d\Omega \frac{C''\left(\Omega\right)}{\Omega^2} \bigg(\Omega\left[\tau_1+\tau_3\right]+\sin\Omega(\tau_1+\tau_2)+\sin\Omega(\tau_2+\tau_3)\\
    & \qquad -\sin\Omega\tau_1-\sin\Omega\tau_2-\sin\Omega\tau_3 -\sin\Omega(\tau_1+\tau_2+\tau_3) \bigg)  \Bigg) \\
    & - {{\mu}}_{\alpha_1}^{10} {{\mu}}_{\alpha_2}^{01} {{\mu}}_{\alpha_3}^{21} {{\mu}}_{\alpha_4}^{12} 
    \exp\left[-i\omega_{10}\tau_1\right] \exp\left[-i\omega_{21}\tau_3\right]  \\
     & \quad \times \exp \Bigg( -\int_0^{\infty} d\Omega \coth{\left(\frac{\beta\hbar\Omega}{2}\right)} 
    \frac{C''\left(\Omega\right)}{\Omega^2} \bigg(2-\cos\Omega\tau_1-\cos\Omega\tau_2-\cos\Omega\tau_3  \\
    & \qquad +\cos\Omega(\tau_1+\tau_2)+\cos\Omega(\tau_2+\tau_3)-\cos\Omega(\tau_1+\tau_2+\tau_3) \bigg)  \\
    & \quad +i\int_0^{\infty} d\Omega \frac{C''\left(\Omega\right)}{\Omega^2} \bigg(\Omega\tau_1+3\Omega\tau_3-\sin\Omega(\tau_1+\tau_2)  \\
    & \qquad -\sin\Omega(\tau_2+\tau_3)-\sin\Omega\tau_1+\sin\Omega\tau_2-3\sin\Omega\tau_3+\sin\Omega(\tau_1+\tau_2+\tau_3) \bigg)  \Bigg)-c.c.\Bigg].
\end{split}
\end{equation}

\begin{equation}
\begin{split}
    \bm{R}_{R,\alpha_4\alpha_3\alpha_2\alpha_1}^{(3)}(\tau_3,\tau_2,\tau_1) 
    = & \; \left(\frac {i}{\hbar}\right)^3 
    \Bigg[ 2{{\mu}}_{\alpha_1}^{10} {{\mu}}_{\alpha_2}^{01} {{\mu}}_{\alpha_3}^{10} {{\mu}}_{\alpha_4}^{01} 
    \exp\left[-i\omega_{10}(\tau_3-\tau_1)\right] \\
    & \quad \times \exp \Bigg( -\int_0^{\infty} d\Omega \coth{\left(\frac{\beta\hbar\Omega}{2}\right)} 
    \frac{C''\left(\Omega\right)}{\Omega^2} \bigg(2-\cos\Omega\tau_1+\cos\Omega\tau_2-\cos\Omega\tau_3  \\
    & \qquad -\cos\Omega(\tau_1+\tau_2)-\cos\Omega(\tau_2+\tau_3)+\cos\Omega(\tau_1+\tau_2+\tau_3) \bigg)  \\
    & \quad +i\int_0^{\infty} d\Omega \frac{C''\left(\Omega\right)}{\Omega^2} \bigg(\Omega\left[\tau_3-\tau_1\right]+\sin\Omega(\tau_1+\tau_2)-\sin\Omega(\tau_2+\tau_3) \\
    & \qquad -\sin\Omega\tau_1-\sin\Omega\tau_2-\sin\Omega\tau_3 +\sin\Omega(\tau_1+\tau_2+\tau_3) \bigg)  \Bigg) \\
    & - {{\mu}}_{\alpha_1}^{10} {{\mu}}_{\alpha_2}^{01} {{\mu}}_{\alpha_3}^{21} {{\mu}}_{\alpha_4}^{12} 
    \exp\left[i\omega_{10}\tau_1\right] \exp\left[-i\omega_{21}\tau_3\right]  \\
     & \quad \times \exp \Bigg( -\int_0^{\infty} d\Omega \coth{\left(\frac{\beta\hbar\Omega}{2}\right)} 
    \frac{C''\left(\Omega\right)}{\Omega^2} \bigg(2-\cos\Omega\tau_1+\cos\Omega\tau_2-\cos\Omega\tau_3  \\
    & \qquad -\cos\Omega(\tau_1+\tau_2)-\cos\Omega(\tau_2+\tau_3)+\cos\Omega(\tau_1+\tau_2+\tau_3) \bigg)  \\
    & \quad +i\int_0^{\infty} d\Omega \frac{C''\left(\Omega\right)}{\Omega^2} \bigg(3\Omega\tau_3-\Omega\tau_1+\sin\Omega(\tau_1+\tau_2)  \\
    & \qquad -\sin\Omega(\tau_2+\tau_3)-\sin\Omega\tau_1-\sin\Omega\tau_2-3\sin\Omega\tau_3+\sin\Omega(\tau_1+\tau_2+\tau_3) \bigg)  \Bigg)-c.c.\Bigg].
\end{split}
\end{equation}
\end{widetext}
First-order perturbation to harmonic energy levels followed by insertion of dipole matrix elements leads to the final expression of quantum non-linear response. 
\begin{widetext}
\begin{equation}
\begin{aligned}
    \bm{R}_{NR,\alpha_4\alpha_3\alpha_2\alpha_1}^{(3)}(\tau_3,\tau_2,\tau_1) 
    = & \; -\frac {i}{2\hbar\omega_0^{2}} 
    \Bigg[ {{\mu}}_{\alpha_1}^{1} {{\mu}}_{\alpha_2}^{1} {{\mu}}_{\alpha_3}^{1} {{\mu}}_{\alpha_4}^{1} 
    \exp\left[-i\omega_{0}(\tau_3+\tau_1)\right] \left(1-2i\hbar\Delta(\tau_3+\tau_1)\right) \\
    & \quad \times \exp \Bigg( -\int_0^{\infty} d\Omega \coth{\left(\frac{\beta\hbar\Omega}{2}\right)} 
    \frac{C''\left(\Omega\right)}{\Omega^2} \bigg(2-\cos\Omega\tau_1-\cos\Omega\tau_2-\cos\Omega\tau_3   \\
    & \qquad +\cos\Omega(\tau_1+\tau_2)+\cos\Omega(\tau_2+\tau_3)-\cos\Omega(\tau_1+\tau_2+\tau_3) \bigg)  \\
    & \quad +i\int_0^{\infty} d\Omega \frac{C''\left(\Omega\right)}{\Omega^2} \bigg(\Omega(\tau_1+\tau_3)+\sin\Omega(\tau_1+\tau_2)+\sin\Omega(\tau_2+\tau_3)  \\
    & \qquad -\sin\Omega\tau_1-\sin\Omega\tau_2-\sin\Omega\tau_3 -\sin\Omega(\tau_1+\tau_2+\tau_3) \bigg)  \Bigg) \\
    & - {{\mu}}_{\alpha_1}^{1} {{\mu}}_{\alpha_2}^{1} {{\mu}}_{\alpha_3}^{1} {{\mu}}_{\alpha_4}^{1} 
    \exp\left[-i\omega_{0}(\tau_3+\tau_1)\right] \left(1-2i\hbar\Delta(\tau_1+2\tau_3)\right) \\
    & \quad \times \exp \Bigg( -\int_0^{\infty} d\Omega \coth{\left(\frac{\beta\hbar\Omega}{2}\right)} 
    \frac{C''\left(\Omega\right)}{\Omega^2} \bigg(2-\cos\Omega\tau_1-\cos\Omega\tau_2-\cos\Omega\tau_3  \\
    & \qquad +\cos\Omega(\tau_1+\tau_2)+\cos\Omega(\tau_2+\tau_3)-\cos\Omega(\tau_1+\tau_2+\tau_3) \bigg) \\
    & \quad +i\int_0^{\infty} d\Omega \frac{C''\left(\Omega\right)}{\Omega^2} \bigg(\Omega\tau_1+3\Omega\tau_3-\sin\Omega(\tau_1+\tau_2)  \\
    & \qquad -\sin\Omega(\tau_2+\tau_3)-\sin\Omega\tau_1+\sin\Omega\tau_2-3\sin\Omega\tau_3+\sin\Omega(\tau_1+\tau_2+\tau_3) \bigg)  \Bigg)\Bigg]
\end{aligned}
\end{equation}

\begin{equation}
\begin{aligned}
    \bm{R}_{R,\alpha_4\alpha_3\alpha_2\alpha_1}^{(3)}(\tau_3,\tau_2,\tau_1) 
    = & \; \left(-\frac {i}{2\hbar\omega_0^{2}}\right) 
    \Bigg[ {{\mu}}_{\alpha_1}^{1} {{\mu}}_{\alpha_2}^{1} {{\mu}}_{\alpha_3}^{1} {{\mu}}_{\alpha_4}^{1} 
    \exp\left[-i\omega_{0}(\tau_3-\tau_1)\right]\left(1-2i\hbar\Delta(\tau_3-\tau_1)\right) \\
    & \quad \times \exp \Bigg( -\int_0^{\infty} d\Omega \coth{\left(\frac{\beta\hbar\Omega}{2}\right)} 
    \frac{C''\left(\Omega\right)}{\Omega^2} \bigg(2-\cos\Omega\tau_1+\cos\Omega\tau_2-\cos\Omega\tau_3 \\
    & \qquad -\cos\Omega(\tau_1+\tau_2)-\cos\Omega(\tau_2+\tau_3)+\cos\Omega(\tau_1+\tau_2+\tau_3) \bigg) \\
    & \quad +i\int_0^{\infty} d\Omega \frac{C''\left(\Omega\right)}{\Omega^2} \bigg(\Omega(\tau_3-\tau_1)+\sin\Omega(\tau_1+\tau_2)-\sin\Omega(\tau_2+\tau_3) \\
    & \qquad -\sin\Omega\tau_1-\sin\Omega\tau_2-\sin\Omega\tau_3 +\sin\Omega(\tau_1+\tau_2+\tau_3) \bigg)  \Bigg) \\
    & - {{\mu}}_{\alpha_1}^{1} {{\mu}}_{\alpha_2}^{1} {{\mu}}_{\alpha_3}^{1} {{\mu}}_{\alpha_4}^{1} 
    \exp\left[-i\omega_{0}(\tau_3-\tau_1)\right]\left(1-2i\hbar\Delta(2\tau_3-\tau_1)\right) \\
    & \quad \times \exp \Bigg( -\int_0^{\infty} d\Omega \coth{\left(\frac{\beta\hbar\Omega}{2}\right)} 
    \frac{C''\left(\Omega\right)}{\Omega^2} \bigg(2-\cos\Omega\tau_1+\cos\Omega\tau_2-\cos\Omega\tau_3 \\
    & \qquad -\cos\Omega(\tau_1+\tau_2)-\cos\Omega(\tau_2+\tau_3)+\cos\Omega(\tau_1+\tau_2+\tau_3) \bigg) \\
    & \quad +i\int_0^{\infty} d\Omega \frac{C''\left(\Omega\right)}{\Omega^2} \bigg(3\Omega\tau_3-\Omega\tau_1+\sin\Omega(\tau_1+\tau_2)  \\
    & \qquad -\sin\Omega(\tau_2+\tau_3)-\sin\Omega\tau_1-\sin\Omega\tau_2-3\sin\Omega\tau_3+\sin\Omega(\tau_1+\tau_2+\tau_3) \bigg)  \Bigg)\Bigg]
\end{aligned}
\end{equation}
\end{widetext}

\bibliography{references}
\end{document}